\documentclass[twocolumn]{aastex631}

\begin{document}

\title{Linear and circular polarimetry of the optically bright relativistic Tidal Disruption Event AT~2022cmc}

\author[0000-0001-7101-9831]{Aleksandar Cikota}
\affiliation{Gemini Observatory / NSF's NOIRLab, Casilla 603, La Serena, Chile}
\affiliation{European Organisation for Astronomical Research in the Southern Hemisphere (ESO), Alonso de Cordova 3107, Vitacura, Casilla 19001, Santiago de Chile, Chile}

\author[0000-0002-8597-0756]{Giorgos Leloudas}
\affiliation{DTU Space, National Space Institute, Technical University of Denmark, Elektrovej 327, DK-2800 Kgs. Lyngby, Denmark}

\author[0000-0002-8255-5127]{Mattia Bulla}
\affiliation{Department of Physics and Earth Science, University of Ferrara, via Saragat 1, I-44122 Ferrara, Italy}
\affiliation{INFN, Sezione di Ferrara, via Saragat 1, I-44122 Ferrara, Italy}
\affiliation{INAF, Osservatorio Astronomico d’Abruzzo, via Mentore Maggini snc, 64100 Teramo, Italy}

\author[0000-0002-9589-5235]{Lixin Dai}
\affiliation{Department of Physics, The University of Hong Kong, Pokfulam Road, Hong Kong}

\author[0000-0003-0733-7215]{Justyn Maund}
\affiliation{Department of Physics and Astronomy, The University of Sheffield, Hicks Building, Hounsfield Road, Sheffield, S3 7RH, UK}

\author[0000-0002-8977-1498]{Igor Andreoni}
\altaffiliation{Neil Gehrels Fellow}
\affiliation{Joint Space-Science Institute, University of Maryland, College Park, MD 20742, USA}
\affiliation{Department of Astronomy, University of Maryland, College Park, MD 20742, USA}
\affiliation{Astrophysics Science Division, NASA Goddard Space Flight Center, Mail Code 661, Greenbelt, MD 20771, USA}

\begin{abstract}

Tidal disruption events (TDEs) occur when a star orbiting a massive black hole is  sufficiently close to be tidally ripped apart by the black hole. AT 2022cmc is the first relativistic TDE that was observed (and discovered) as an optically bright and fast transient, showing signatures of non-thermal radiation induced by a jet which is oriented towards the Earth.
In this work, we present optical linear and circular polarization measurements, observed with VLT/FORS2 in the $R$-band (which corresponds to the blue/UV part of the spectrum in rest frame), $\sim$ 7.2 and $\sim$ 12.2 rest-frame days after the first detection, respectively, when the light curve of the transient had settled in a bright blue plateau. 
Both linear and circular polarization are consistent with zero, $p_{lin}$ = 0.14 $\pm$ 0.73 \% and $p_{cir}$ = $-$0.30 $\pm$ 0.53\%. 
This is the highest S/N linear polarization measurement obtained for a relativistic TDE and the first circular polarimetry for such a transient.
The non detection of the linear and circular polarization is consistent with the scenario of AT 2022cmc being a TDE where the thermal component (disk+outflows) is viewed pole-on, assuming an axially symmetric geometry. 
The presence and effect of a jet and/or external shocks are, however, difficult to disentangle.
\end{abstract}

\section{Introduction}

Tidal disruptions of stars around supermassive black holes \citep[][]{Rees1988,Phinney1989} have been discovered using a variety of methods at different wavelengths \citep[][]{komossa99,Gezari2006,Levan2011,Gezari2012}. Nowadays, the majority of TDEs are found by wide-field optical surveys and a distinct class of objects has been discovered where the optical/UV emission is dominated by a thermal component with black-body temperatures $2 - 4 \times 10^4$~K \citep{vanvelzen2020}. 
While these objects, primarily found in quiescent or post-starburst galaxies \citep{Arcavi2014,french2020ssrv}, constitute the most numerous and homogeneous class among TDEs, there exist a number of other transients that have been proposed to be due to tidal disruptions. These include objects found through their X-ray emission \citep{2020SSRv..216...85S}, objects found in Active galactic nuclei (AGNs) \citep[explicitly excluded by the criteria of][]{vanvelzen2020}, such as PS16dtm \citep{BlanchardPS16dtm}, objects found in luminous infrared galaxies often suffering considerable extinction \citep{Tadhunter2017,Mattila2018}, unique objects such as the superluminous ASASSN-15lh \citep{Leloudas2016}, some of which are often debated to be of supernova or AGN origin \citep[see][]{Zabludoff2021}, or fast blue optical transients (FBOTs), such as AT~2018cow, whose nature is also  debated  \citep[e.g.][]{2019MNRAS.484.1031P, 2019MNRAS.487.2505K, 2021ApJ...911..104K, 2022ApJ...932...84M, 2022ApJ...935L..34L}.

Within the TDE zoo, there also exist a small number of objects that have been discovered through their high-energy emission, by triggering the Burst Alert Telescope (BAT) on the Neil Gehrels \textit{Swift} Observatory. 
These include Swift J164449.3+57345 \citep{Bloom2011,Levan2011,Zauderer2011}, 
Swift J2058.4+0516 \citep{Cenko2012,2015ApJ...805...68P}
and Swift J1112.2-8238 \citep{Brown2015}.
These transients, found typically at higher redshifts (up to $z \sim 1.2$), exhibit strong and long-lasting X-ray emission with rapid variability (differentiating them from Gamma-ray burst afterglows) accompanied by radio emission, attributed to synchrotron radiation.  
The consensus is that these objects are TDEs that are thought to have  relativistic jets, and, for this reason, are also known as relativistic or ``jetted" TDEs.

AT~2022cmc is another transient at $z = 1.193$ \citep{2022GCN.31602....1T} that has been proposed to be a relativistic TDE \citep{Andreoni2022,Pasham2022}, although it was not discovered through its high-energy emission. It was discovered in the optical by a wide-field transient survey, the Zwicky Transient Facility \cite[ZTF;][]{BellmZTF}, where it was initially noticed as a fast and red transient \citep{2022TNSAN..38....1A}. 
\citet{Andreoni2022} presented a detailed study of this event for the first 15 rest-frame days, including X-ray, UV/optical/IR and sub-millimeter/radio data, suggesting that a relativistic jet was formed producing an afterglow powered by synchrotron radiation. In the UV/optical wavelengths, they showed that the initial rapid (and red) decay phase is followed by a bluer luminous ($M_g \sim -22$  mag) plateau $\sim$4.5 days after detection in the rest frame ( $\sim$10 days in observer frame, see their Fig. 1) and they argue that this second phase traces a more ordinary thermal component, reminiscent of optical TDEs.
\citet{Andreoni2022} examined alternative solutions for the nature of this transient, including a kilonova, FBOT, and a Gamma-ray burst (GRB) origin, but they reject all of them supporting the scenario of a relativistic TDE.
This conclusion is corroborated by \citet{Pasham2022} who show that the X-ray emission of AT~2022cmc demonstrates rapid variability on timescales of hours, requiring a small emitting region, seen previously in relativistic TDEs but not in GRB afterglows. Their spectral energy distribution (SED) modeling show that the X-rays are more likely produced through synchrotron self-Compton (SSC) emission, while radio emission is standard synchrotron emission. A clear thermal component dominates in the optical/UV at $\sim$ 11 days in the rest frame ($\sim$ 25 days in the observer frame, see their Fig. 3).

Here we present optical (rest-frame UV) linear and circular polarimetry of AT~2022cmc obtained at 15.84 and 26.81 observer-frame days (7.22 and 12.23 rest-frame days) respectively, relative to the first detection on February 11, 2022 at 10:42:40 UTC (MJD 59621.44), following \citet{Andreoni2022}. Thus, the blue plateau dominates the transient evolution during our observed epochs.

In Sect.~\ref{sect:Obs} we explain the observations, in Sect.~\ref{sect:MethodsResults} we present the methods and results, and in Sect.~\ref{sect:discussion} we discuss our results in the context of the proposed scenarios for this transient.
Section~\ref{sect:conc} contains our summary and concluding remarks.

\section{Observations}
\label{sect:Obs}
We obtained imaging linear and circular polarimetry of AT 2022cmc ($\alpha$=13:34:43.207, $\delta$=+33:13:00.54) on February 27, 2022 at 06:57h UT (MJD 59637.29) and March 10, 2022 at 06:16h UT (MJD 59648.26), respectively, with the FOcal Reducer/low dispersion Spectrograph 2 (\citealt{1998Msngr..94....1A}, FORS2) mounted on the primary focus of European Southern Observatory's (ESO) Very Large Telescope (VLT) Antu (UT1). 
All observations were obtained using the FORS2 R\_SPECIAL filter ($\lambda_0$ = 655 nm, FWHM = 165 nm), which corresponds to the central wavelength of $\lambda_0$ = 299 nm at $z$ = 1.193. 
The brightness of AT 2022cmc at these phases was 21.8 $\pm$ 0.4 $R$ mag 
and 22.0 $\pm$ 0.4 $R$ mag, 
measured with aperture photometry in the FORS2 acquisition images using the Vega photometric system.

\begin{figure*}
\centering
\includegraphics[trim=0mm 0mm 0mm 0mm, width=8.8 cm, clip=true]{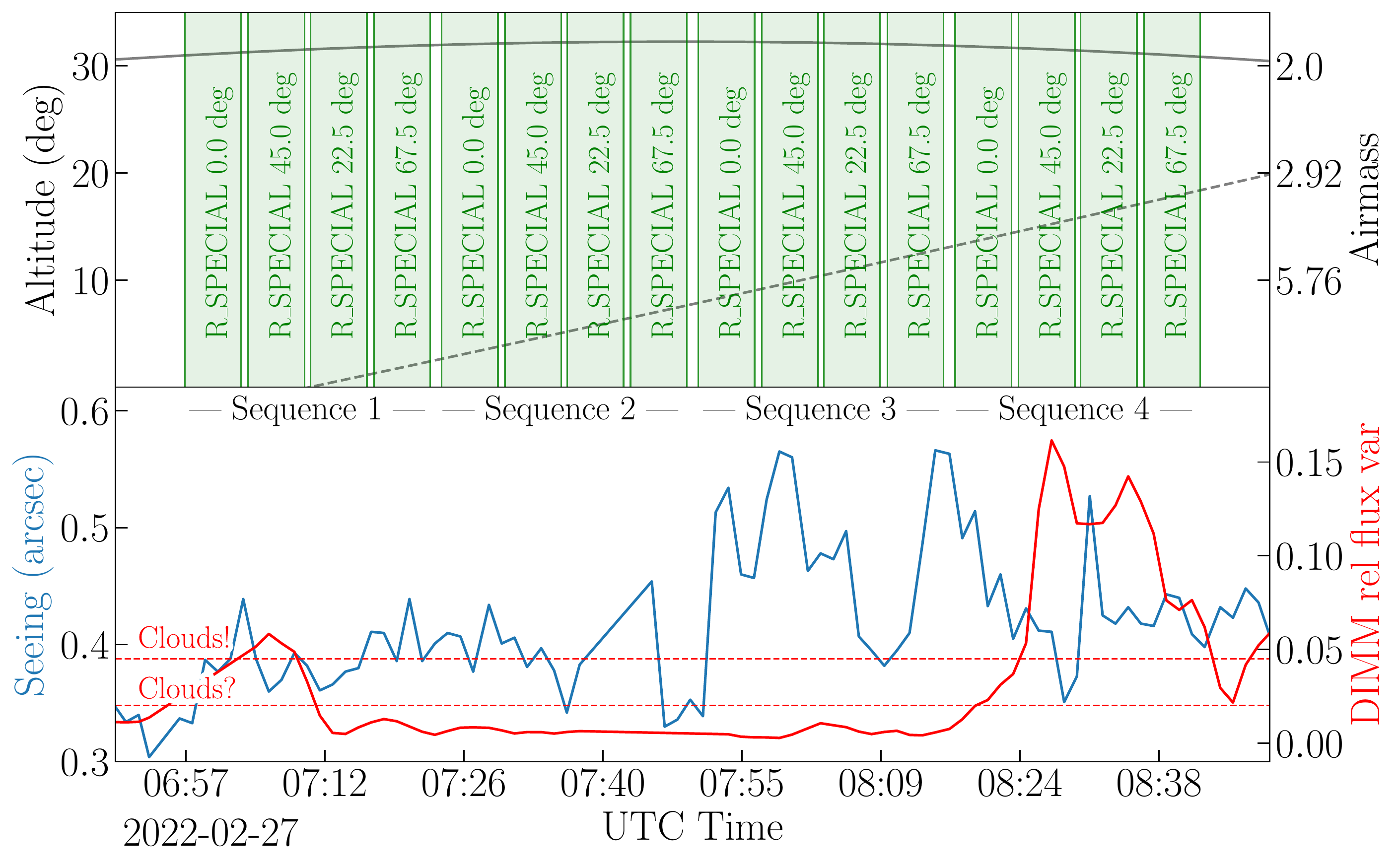}
\hspace{1mm}
\includegraphics[trim=0mm 0mm 0mm 0mm, width=8.8 cm, clip=true]{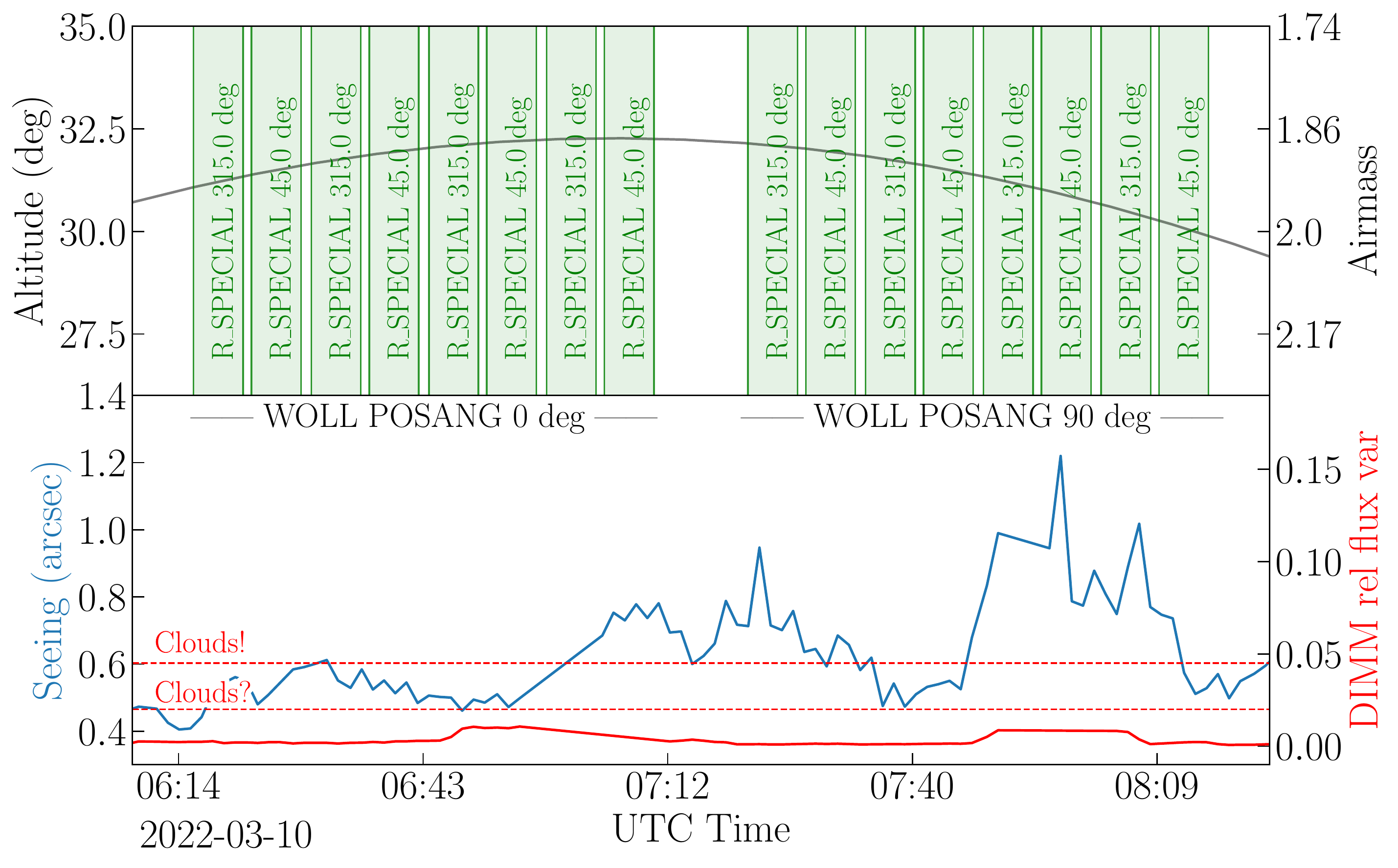}
\vspace{-2mm}
\caption{Observing log of linear and circular polarimetry of AT 2022cmc. The green blocks on the top panels indicate the individual exposures, each of 350 sec in the $R$-band, and the corresponding HWP angles. The solid and dashed lines show the altitudes of the target and Moon, respectively. 
The bottom panels show the seeing measured by the ASM-DIMM telescope at 500 nm (blue line), and the relative flux variation along the line of sight of the DIMM telescope (red line) during the observations. The dashed red lines indicate thresholds for the possible presence of clouds. 
\textit{Left:} The linear polarimetry sequence of 4 HWP angles was repeated 4 times to increase the S/N. The individual sequences are indicated. \textit{Right:} Circular polarimetry was obtained at two orientations of the instrument, separated by 90 degrees. At each orientation 4 sequences of two quarter-wave plate positions were taken.}
\label{fig:ObsLog}
\end{figure*}

Figure~\ref{fig:ObsLog} shows the observing log with individual exposures for the linear polarization (left panel) and circular polarization observations (right panel). The seeing conditions, measured with the Differential Image Motion Monitor (\citealt{1990A&A...227..294S}, DIMM), have been downloaded from the Ambient Conditions Database\footnote{\href{http://archive.eso.org/cms/eso-data/ambient-conditions.html}{http://archive.eso.org/cms/eso-data/ambient-conditions.html}}.

Linear polarimetry was acquired using four half-wave plate (HWP) angles of 0, 22.5, 45, and 67.5 degrees, and repeated four times to increase the S/N. The redundancy of four HWP angles reduces the flat-fielding issue and cancels out other instrumental effects (see \citealt{Patat2006}). AT 2022cmc was at its peak altitude during the observations, between $\sim$30 and $\sim$33 degrees, at an airmass of just above 2. The seeing was stable during the observations, between $\sim$ 0.3 and 0.6", and the sky transparency determined by the weather officer was clear, however, during the fourth sequence, there were some thin clouds passing (see Fig.~\ref{fig:ObsLog}). The measured FWHM of the target was between $\sim$0.7" and 0.9" during the observations. The Moon, at a distance of $\sim$ 105 degree from AT 2022cmc and on illumination of 15\%, started rising at $\sim$ 7 am UT and reached an altitude of $\sim$ 20 degree by the end of the observations.

Circular polarimetry was obtained with two different quarter-wave retarder plate (QWP) angles of $\pm$45 degrees. The sequence of two QWP angles has been repeated four times to increase the S/N. Furthermore, the set of 4x2 angles have been taken at two different rotations of the instrument of 0 and 90 degrees in order to eliminate possible crosstalk between linear and circular polarization \citep[see][]{2009PASP..121..993B}. 
The target was again observed during its peak altitude, at an airmass of $\lesssim$2 and with the Moon set. The sky transparency was clear during the execution of the observing block, however, the seeing was variable and increased up to $\sim$1.2" during the second set of observations with the instrument rotated. The measured FWHM of the target was between $\sim$0.9" and 1.2" during the observations.

\section{Methods and results}
\label{sect:MethodsResults}

\subsection{Linear polarimetry}
\label{subsection:linearpolarimetry}

To calculate the linear polarization of AT 2022cmc, we measured the flux of the target in the ordinary and extraordinary beams at all HWP angles using the aperture photometry function from PythonPhot\footnote{\href{https://github.com/djones1040/PythonPhot}{https://github.com/djones1040/PythonPhot}}. We used an aperture radius of 1.3", with the inner and outer sky radii of 2.5" and 5", respectively. Tests with slight variations of the aperture and sky radii produced consistent results.
The aperture sizes and positions were fixed for all images within the sequences.

The normalized Stokes $q$ and $u$ parameters were calculated following the standard approach as described in the FORS2 User Manual (\citealt{FORS2manual}, see also \citealt{2017MNRAS.464.4146C}, and \citealt{2022MNRAS.509.6028C}):
\begin{equation}
\label{eq:stokes}
\begin{array}{l}
q = \frac{2}{N} \sum_{i=0}^{N-1} F (\theta_i) \cos (4\theta_i), \\ 
u = \frac{2}{N} \sum_{i=0}^{N-1} F (\theta_i) \sin (4\theta_i),
\end{array}
\end{equation}
where $N$ ranges over the half-wave retarder plate angles, $\theta_i = 22.5 ^\circ \times i$, with $0 \leq i \leq 3$, and $F(\theta_i)$ is the normalised flux difference between the ordinary ($f^o$) and extraordinary ($f^e$) beams:

\begin{equation}
\label{eq:normfluxdiff}
F (\theta_i) = \frac{f^o (\theta_i) - f^e (\theta_i)}{f^o 
(\theta_i) + f^e (\theta_i)}.
\end{equation}

The polarization position angles of the raw measurements have been corrected for the half-wave plate zero angle chromatic dependence \citep[Table 4.7 in][]{FORS2manual}.

The reduction pipeline and possible instrumental effects (i.e. instrumental polarization and angle offset) have been verified using archival observations of polarized and unpolarized standard stars. 

The measured Stokes parameters for AT 2022cmc are $q$ =  0.20 $\pm$ 0.67 \% and $u$ = 0.17 $\pm$ 0.79 \%, which leads to a polarization degree of $p_{lin}$ = 0.26 $\pm$ 0.73 \%, and after a polarization bias correction, following \citet{2014MNRAS.439.4048P}, $p_{lin_{BiasCorr}}$ = 0.14 $\pm$ 0.73 \%, i.e. a 3$\sigma$ upper limit of $\sim$ 2.3\%. We note that the host galaxy of AT2022cmc is not detected to very faint limits ($>24.5$ mag; \citealt{Andreoni2022}) and it can therefore contribute maximum $6$\% of the captured light. For this reason we have not applied any host galaxy dilution correction for the transient polarization \citep{Leloudas2022,2022arXiv220814465L}.

\begin{figure}
	\includegraphics[width=\columnwidth]{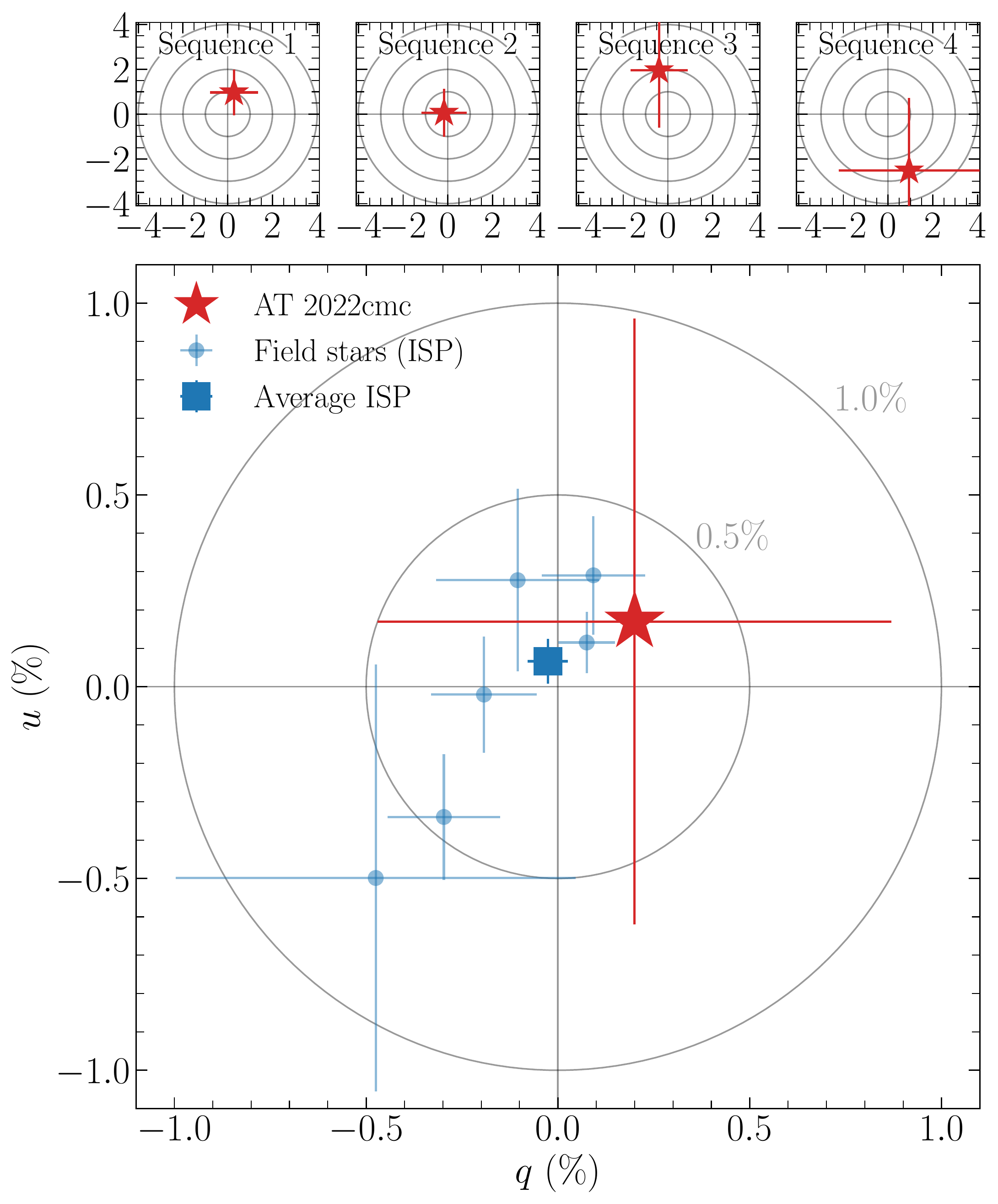}
    \caption{Linear polarization of AT 2022cmc (red star) in the Stokes $q$-$u$ plane at 7.22 rest-frame days after first detection. The light blue dots display the interstellar polarization (ISP) measured through field stars and the blue square is their weighted mean. The polarization of AT 2022cmc measured in the individual sequences is displayed in the four subplots on the top of the figure. The gray circles in the subplots denote polarization degrees of 1, 2, 3 and 4 per cent. The polarization of AT 2022cmc is consistent with zero and this is also true for the ISP.}
    \label{fig:linpol_results}
\end{figure}

The polarization of AT 2022cmc is shown on the Stokes $q$--$u$ plane in Fig.~\ref{fig:linpol_results}. The main plot shows the polarization determined from all data combined, i.e. by calculating the normalized flux difference for each HWP angle using all 4 sequences of the 4 HWP angles before calculating the Stokes $q$ and $u$.

The small subplots at the top of Fig.~\ref{fig:linpol_results} display the polarization of AT 2022cmc measured using the individual sequences (which consists of 4 HWP angles). All the measurements are consistent with zero polarization. The measurement in Sequence 4 has been affected by thin clouds passing (see Fig.~\ref{fig:ObsLog}) and therefore exhibits larger error bars and a larger offset compared to the first 3 sequences. If we exclude the last sequence and determine the polarization from first three sequences only, the result remains consistent within the uncertainties.

In addition, we used 6 bright field stars to determine the interstellar polarization (ISP) by calculating their weighted mean in the $q$--$u$ plane (Fig.~\ref{fig:linpol_results}). We corrected the field stars polarization measurements for the instrumental polarization (\citealt{Patat2006,2020A&A...634A..70G}) which increases with distance from the optical axis using the same methods as applied in \citet[][see their Fig.~2]{2022MNRAS.509.6028C} and \citealt{2015ApJ...815L..10L}. The mean ISP determined from field stars is negligible, $p_{lin}$ = 0.03 $\pm$ 0.06 \%. 
This is consistent with the low values found for nearby stars in the catalogue by \cite{2000AJ....119..923H} and with the low Galactic reddening, $E(B-V)$ = 0.0095 $\pm$ 0.0006 mag, in the direction of AT 2022cmc  \citep{2011ApJ...737..103S}.

\citet{Andreoni2022} suggest that AT~2022cmc does not suffer from any significant host extinction. Therefore, we also do not expect significant host galaxy ISP, which is also consistent with the non-detection of linear polarization.

\subsection{Circular polarimetry}

The flux of AT 2022cmc was extracted from the ordinary and extra-ordinary beams using the same tools as described in the previous section (Sect.~\ref{subsection:linearpolarimetry}), and the amount of circular polarization was calculated following the equation for the Stokes $v$ given in the FORS2 user manual \citep{FORS2manual}:
\begin{equation}
{v} =\frac{1}{2}  \left[ \left(\frac{f^o - f^e}{f^o + f^e}\right)_{\theta=45^{\circ}}    -  \left(\frac{f^o - f^e}{f^o + f^e}\right)_{\theta=-45^{\circ}}  \right]
\end{equation}
where $f^o$, and $f^e$ is the flux measured in the ordinary and extra-ordinary beams, respectively, for both quarter-wave retarder plate angles of $\theta= \pm 45^{\circ}$.
For circular polarimetry, we used an aperture radius of 1.5", with the inner and outer sky radii of 2.5" and 5", respectively.

\begin{figure}
	\includegraphics[width=\columnwidth]{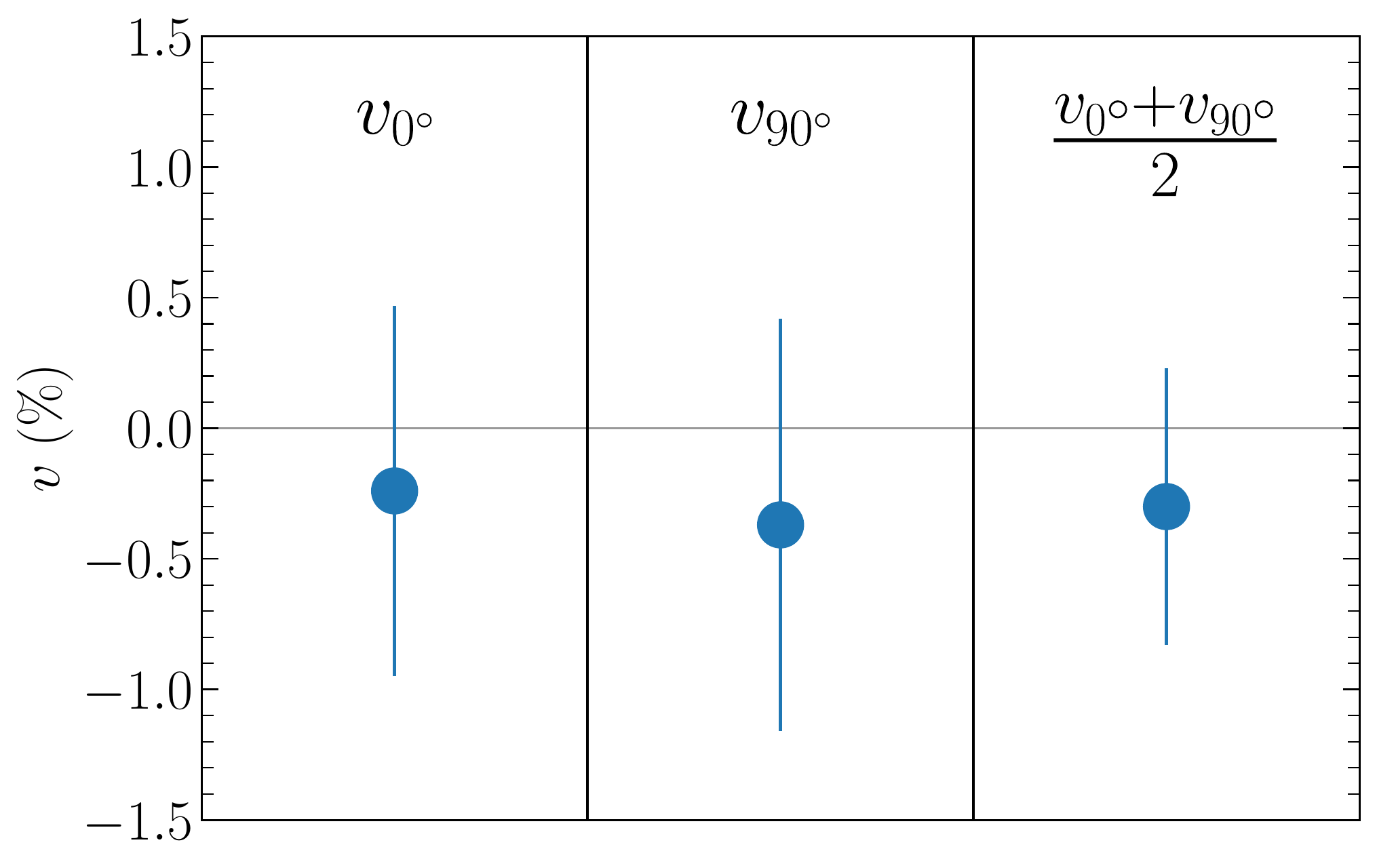}
    \caption{Circular polarization of AT 2022cmc at 12.23 days after first detection in the rest-frame. The left and middle panel show the circular polarization measured at two different orientations of the instrument, separated by 90 degrees, and the right panel shows the average of both measurements. The circular polarization is consistent with zero. }
    \label{fig:circpol_results}
\end{figure}

Figure~\ref{fig:circpol_results} displays the circular polarization measured at the two different orientations of the instrument separated by 90 degrees (left and middle panels). The two measurements of $v_{0^{\circ}}$ = -0.24 $\pm$ 0.71~\% (observed with the instrument aligned toward the North Celestial Pole) and 
$v_{90^{\circ}}$ = -0.37 $\pm$ 0.79 \% (observed with the instrument rotated by 90 degrees) are consistent. Furthermore, the weighted average of the two measurements, which benefits from the cancellation of the possible spurious signal (introduced because of linear-to-circular polarization cross talk, see \citealt{2009PASP..121..993B}) is 
$p_{cir}$ = -0.30 $\pm$ 0.53\% (right panel in Fig.~\ref{fig:circpol_results}).
Thus, we did not detect significant circular polarization for AT 2022cmc.
Note that in the case of circular polarimetry no bias correction is needed, in contrast to linear polarimetry.

\section{Discussion}
\label{sect:discussion}

Optical polarimetry has been extensively used in order to probe the geometry and physics of transients. 
The origin and degree of polarization varies for different transients and physical mechanisms.
In supernova explosions the polarization is primarily produced due to electron scattering (i.e. Thomson scattering) and the degree of the continuum polarization is related to the degree of asymmetry of the  photosphere \citep{Hoeflich1991,Kasen2003}. Continuum polarization levels range from $<0.3$\% for SNe Ia, indicating they are nearly spherical \citep{2019MNRAS.490..578C}, and up to 1-2\% for core-collapse SNe, with stripped SNe Ib/c being the most asymmetric \citep{WangWheeler,PatatSNhandbook}.
Very few circular polarimetry measurements have been obtained and observations of two SLSNe and a SN Ia only resulted in non detections \citep{Cikota2018, 2013MNRAS.433L..20M}. Despite the very different opacities and their critical role, electron scattering is also expected to be the source of polarization in kilonovae from neutron star mergers \citep{BullaKN, Bulla2021} and AT~2017gfo, related to GW 170817, showed very low polarization levels likely due to ISP \citep[$0.5$\%; ][]{Covino2017}. 
%
%

For GRBs on the other hand, there are several mechanisms that could cause polarization \citep{CovinoGotz}. Although in the first few minutes linear polarization values of up to 30\% induced by reverse shocks have been reported \citep{Mundell2013}, values between 0-3\% have been observed on timescales of hours and days after several GRBs, showing moderate variability with a generally decreasing trend.
The afterglow polarization is attributed to synchrotron emission from  shock-accelerated electrons in the ambient medium (forward shock). 
The measured polarization is strongly dependent on the geometric configuration, whether the viewing angle is within the jet opening angle, the structure of the jet and the nature (ordered versus random) of the magnetic fields \citep[see e.g.][]{2006NJPh....8..131L, 2020ApJ...892..131S, 2021MNRAS.507.5340T, 2004ASPC..312..460R}. The polarization is expected to peak at the jet break time.
Circular polarimetry of GRBs has mostly yielded upper limits, with the exception of a 0.61 $\pm$ 0.13 \% detection for GRB 121024A \citep{Wiersema2014}. Also in the case of blazars 3C~279 and PKS~1510-089, which show variable and high levels of linear polarization in the optical wavelengths (between 0 and $>$30$\%$), no obvious circular polarization was  detected ($p_{cir}<$1$\%$, \citealt{Liodakis2022}).

Polarimetry of TDEs is a nascent field. Interestingly, some of the first studies were made for the rare class of relativistic TDEs. 
\cite{Wiersema2012} measured 7.4 $\pm$ 3.5 \% in the $K_s$-band for Swift J164449.3+57345 at 12.2 rest-frame days after trigger (note that the time of trigger approximately coincides to peak brightness for fast transients). This event was however highly extincted in the optical, and it is not clear how much polarization induced by dust affects this result. Nevertheless, the authors seem to disfavor the contribution from a Synchrotron Self-Compton (SSC) component in the $K$-band polarization, even if the presence of such a component is suggested by SED modeling \citep{Bloom2011}.
Furthermore, \cite{Wiersema2020} presented optical linear polarimetry of Swift J2058.4+0516, which did not suffer from significant extinction. They report $p_{lin}<$5.3\% and $p_{lin}$ = 8.1 $\pm$ 2.5 \% at 40 and 75 days after trigger in the rest-frame, respectively. We note that these were challenging measurements, even for VLT, as the transient was 23--24 mag at the time of observations, which is reflected in the low S/N. 
This polarization could originate from synchrotron radiation from the forward shock in the jet, although these values are larger than what is seen for GRBs at these phases. 
Another exceptionally luminous transient (likely a TDE as argued in \citealt{Leloudas2016}, although a superluminous supernova origin has also been proposed by \citealt{2016Sci...351..257D}) with a good polarimetric coverage is ASASSN-15lh \citep{Maund2020}. This object shows an overall constant polarization level of $\sim0.4$\% between $+$30 and $+$90 days relative to peak brightness in the rest-frame with a single individual measurement showing $p_{lin}$ = 1.2 $\pm$ 0.2 \% near the minimum of the UV light curve, indicating possible rapid variability in timescales of $\pm10$ days. Spectral polarimetry indicates a polarization level that is overall independent of wavelength. 

Polarimetric observations of `ordinary' optical TDEs \citep{vanvelzen2020} were scarce until recently including few measurements of individual events
\citep{HigginsSPLOT,Lee2020,Holoien2020}.
\cite{Leloudas2022} gathered spectral polarimetry for a sample of 3 optical TDEs, and demonstrated that the continuum polarization is wavelength independent, while emission lines depolarize the spectrum. 
They suggest that the origin of polarization in optical TDEs is electron scattering, similar to SNe and kilonovae, and they exclude synchrotron radiation and dust scattering as significant contributors. The polarization is observed to decrease with time and 
\cite{Leloudas2022} propose that the data is compatible with the formation and evolution of a super-Eddington accretion disk.
They further model the polarization with the super-Eddington accretion model of \cite{Dai18} and the radiative transfer code \texttt{POSSIS} \citep{BullaPOSSIS}, yielding polarization predictions between 0--6\% for different observing angles, values that are broadly consistent with the observations. 
A simultaneous study \citep{Patra2022} presented spectral polarimetry of AT~2019qiz, showing zero polarization at peak 
but increasing with time, one month later. 
Finally, the study by \citet{Liodakis2022} presented time-varying polarization for AT 2020mot, reaching an extraordinary 25 $\pm$ 4 \% in one epoch. This measurement is very hard to explain by reprocessing models \citep{Leloudas2022,Charalampopoulos2022} and the authors 
favor shock collisions for this TDE.

\begin{figure}
	\includegraphics[width=\columnwidth]{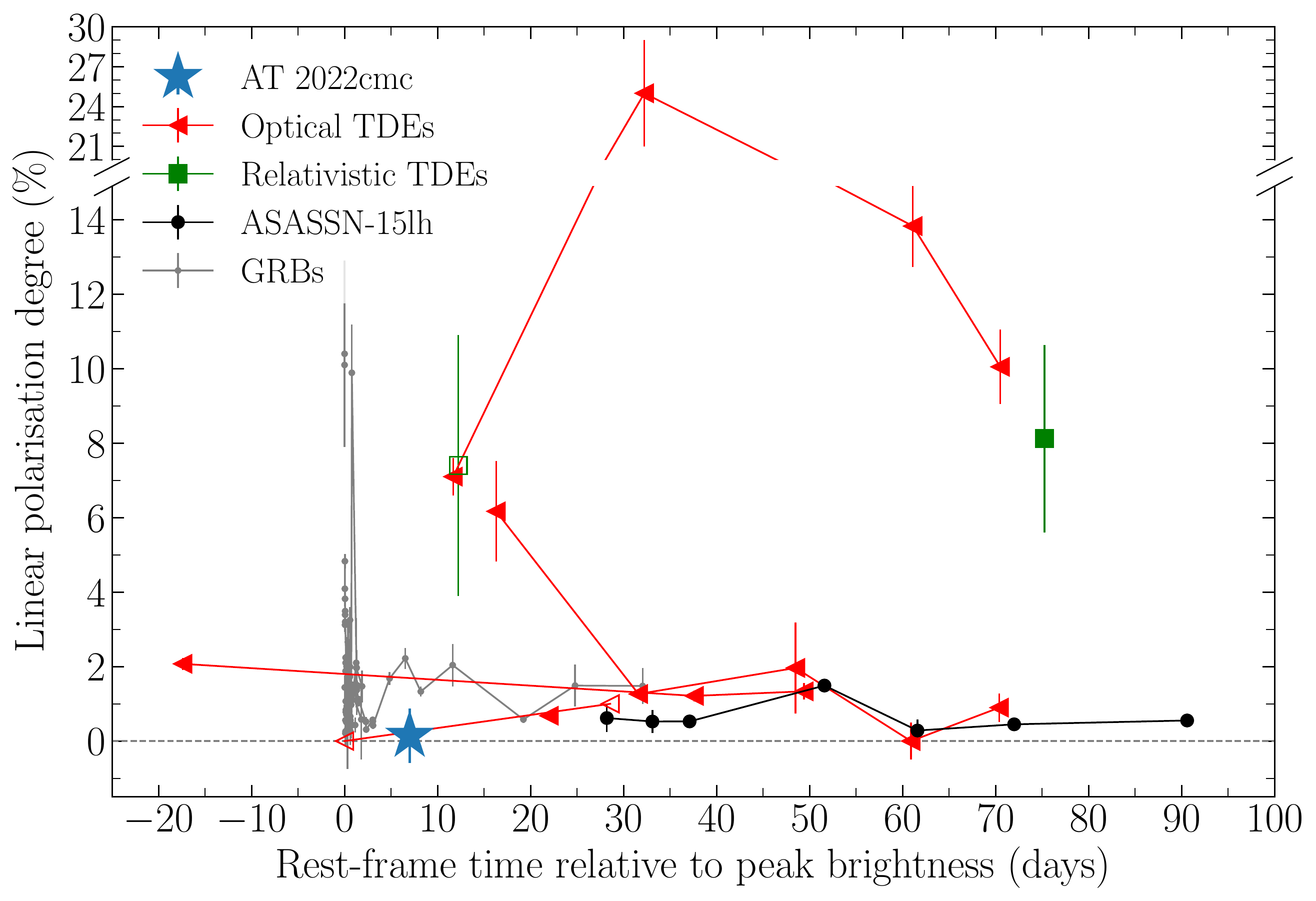}
    \caption{Linear polarization of AT 2022cmc (this work, blue star) compared to the polarization of TDEs AT 2018dyb, AT 2019dsg, AT 2019azh, AT 2019qiz and AT 2020mot (\citealt{Leloudas2022, Patra2022, 2022arXiv220814465L}, red triangles), relativistic TDEs Swift J164449.3+573451 and Swift J2058+0516 (\citealt{Wiersema2012,Wiersema2020}, green squares), the superluminous transient ASASSN-15lh (\citealt{Maund2020}, black dots), and GRBs collected by \citet[][gray dots]{CovinoGotz} with N$>$4 epochs. The polarization is shown as a function of days  relative to peak brightness. Note that the time of first detection approximately coincides to peak brightness for GRBs. TDEs J164449.3+57345 and AT~2019qiz are marked with open symbols to mark uncertain values because they have not been corrected for the host galaxy dilution (see Sect.~\ref{sect:discussion}). }
    \label{fig:linpol_comparison}
\end{figure}

Figure~\ref{fig:linpol_comparison} shows a comparison of the linear polarization of AT~2022cmc with other transients including ordinary optical TDEs  \citep{Leloudas2022, Patra2022, 2022arXiv220814465L}, relativistic  \citep{Wiersema2012,Wiersema2020} and extreme TDEs  \citep[ASASSN-15lh;][]{Maund2020}, and a sample of GRBs selected from \citet[][]{CovinoGotz} with N$>$4 epochs.

We note that the nuclear transients in this figure have either been corrected for host dilution \citep{Leloudas2022,2022arXiv220814465L}\footnote{The data for ASASSN-15lh was not host corrected in \cite{Maund2020} but we applied the correction here for the first time. We find that the host contamination in the $V$-band is small and increases from 13\% to 24\% during the polarimetric observations. This minor correction does not affect the discussion in \cite{Maund2020}.} or such a correction is not necessary, as in the case of Swift J2058+0516 and AT 2022cmc, where the transient was $>$3 mag brighter than the host at the time of linear polarimetry \citep{2015ApJ...805...68P,Andreoni2022}.
Two transients have not been host corrected due to limited or unavailable information and appear in  Figure~\ref{fig:linpol_comparison} with open symbols. These are AT~2019qiz \citep{Patra2022} and Swift J164449.3+57345 \citep{Wiersema2012}. In the first case, the 0\% value at peak will not be affected but the 1\% one month later is a lower limit. The case of Swift J164449.3+57345 is much more complicated as the intrinsic polarization is both uncertain due to the host contribution but also due to considerable contribution by dust \citep{Wiersema2012}. 
The GRB literature data is not host corrected but we expect this correction to not be significant as GRBs are not embedded in their host galaxy nucleus and their afterglows typically outshine the host.

\subsection{The unpolarized case of AT~2022cmc}

The optical/NIR light curve of AT 2022cmc displays a steep decrease in brightness from $\sim$19.0 mag to $\sim$21.5 mag in the $r$-band the first $\sim$5 days (in rest frame), and a plateau after the $\sim$5th day, which lasts at least until the 15th day followed by a slower decrease \citep{Andreoni2022}. 
Both linear and circular polarimetry of AT~2022cmc were obtained during that plateau phase, where the optical light is dominated by a blue component. 

The degree of linear polarization is very low, $p_{lin}$=0.14 $\pm$ 0.73 \%, consistent with zero, and lower than what has been observed in the case of the other two relativistic TDEs, Swift J164449.3+573451 and Swift J2058+0516. It is also lower than the linear polarization measured for 3/5 optical TDEs at 97\% confidence.
The spectrum of AT~2022cmc is featureless (Fig. 3 in \citealt{Andreoni2022}) during the phases of our observations, so a contribution due to depolarizing emission lines \citep{Leloudas2022,Patra2022} can be excluded.

In general, optical linear polarization in TDEs can be produced by at least the following mechanisms: 
(i) by the synchrotron radiation of an on-axis or off-axis jet, as in the case of GRBs or blazars. This has been proposed for previous relativistic TDEs (\citealt{Wiersema2012,Wiersema2020}). Note that in the case of jets we also expect detection of radio and/or X-ray, which was observed in the case of AT~2022cmc \citep{Andreoni2022, Pasham2022};
(ii) stellar stream shocks can produce high polarization degrees ($>$25\%), which arises as the result of multiple competing shocks. This was proposed in the case of AT~2020mot \citep{Liodakis2022};
(iii) Electron scattering from a reprocessing layer in an accretion disk and possible outflows. In the case of the super-Eddington accretion disk model of \citet{Dai18}
the polarization is expected to decrease with the viewing angle, and can reach polarization degrees of up to $\sim$6\% for edge-on disks, depending also on the density distribution of the material in the disk \citep{Leloudas2022}. 
Note that there is a variety of possible TDE accretion models, which may produce different polarization degrees, including the collision-induced outflow \citep{LuBonnerot2020}, which can produce linear polarization up to 9\% \citep{Charalampopoulos2022},
or the zero-Bernoulli accretion flow model \citep{2014ApJ...781...82C}, which assumes that an accretion disc is quasi-spherical, radiation-pressure dominated, and accompanied by the production of strong jets \citep{2022MNRAS.517.6013E}.
Furthermore, clumpy flared disk models may produce higher polarization degrees of up to $\sim$10\% and variable polarization angles \citep{2015sf2a.conf..167M}. Such reprocessed radiation is also expected to be detectable in the radio wavelengths.

It is therefore difficult to confidently exclude any models, based on a single measurement of $p_{lin} < 2.3\%$, especially when this is compatible with zero polarization, which could be realized for specific viewing angles. What is, however, possible is to confirm that 
the zero polarization is consistent with the \texttt{POSSIS} modelling in \cite{Leloudas2022} for a TDE accretion flow viewed relatively close to the pole. This idea is therefore compatible with the scenario proposed by \citet{Andreoni2022}, where at these phases the optical/UV probes a thermal component (outflows) from the TDE. This however assumes an axisymmetric geometry, as in the (idealized) model.

\begin{figure}
	\includegraphics[width=\columnwidth]{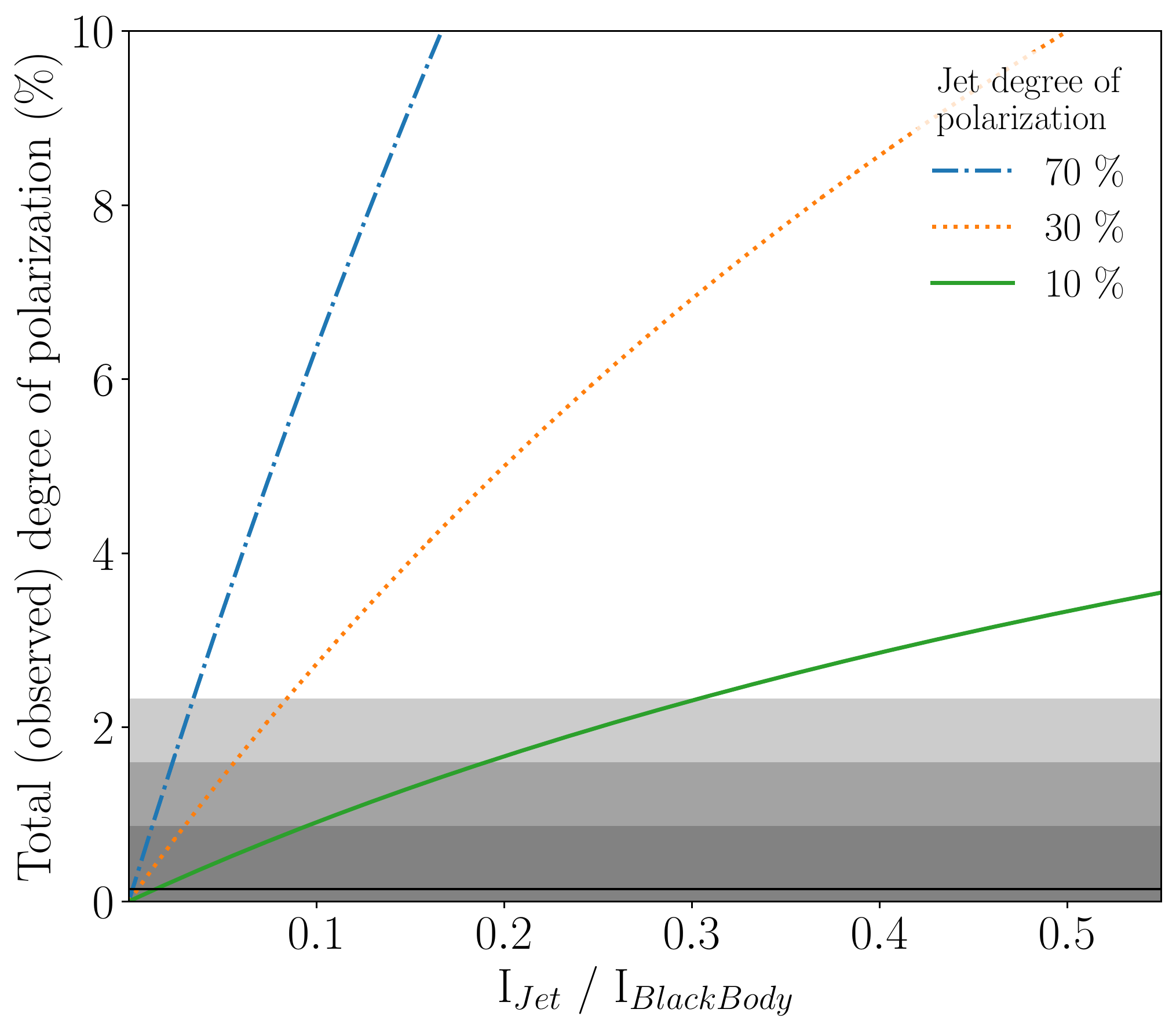}
    \caption{Constraints to the ratio of the non-thermal to the thermal component for AT 2022cmc at 7.2 days, based on our linear polarimetry observations and the degree of polarization of the putative jet. The black line shows the measured linear polarization of 0.14\%, and the gray shaded area indicates the 1, 2 and 3$\sigma$ uncertainties.}
    \label{fig:pol_calc}
\end{figure}

In addition, this ignores any contribution from the jet in the UV/optical already at 7.2 days after discovery. 
This may either mean that the jet component is sub-dominant, or that the jet is not significantly intrinsically polarized.
This is illustrated in 
Figure \ref{fig:pol_calc}, which shows the predicted polarization levels assuming different ratios of the two components and different levels of polarization for the putative jet component, ranging from 10\% to the maximum theoretical value of 70\% for synchrotron radiation (\citealt{2003ApJ...596L..17G, 2003ApJ...594L..83G, 2003JCAP...10..005N}; see also \citealt{CovinoGotz} for a review).
However, that is not the case for GRB afterglows, in which the emission is produced by shocked ambient medium (forward shock) and the intrinsic polarization averages out in the magnetic fields which appear random to the observer. 
Thus, despite the polarization in GRB afterglows is produced due to synchrotron emission, the measured polarisation is strongly dependent on the geometric nature, with only a small contribution from ordered fields. Due to relativistic effects in the observed geometry of the blastwave (i.e. the surface of equal arrival time, see \citealt{2010arXiv1012.5101G}), the detection of small levels of polarisation in afterglows seen on-axis (i.e. the viewing angle is within the jet opening angle) is not surprising (see Fig. 3 in \citealt{2006NJPh....8..131L}; see also \citealt{2003ApJ...595L..25M, 2004ApJ...615..366S, 2008ApJ...673L.123T, 2010A&A...520L...7H, Liodakis2022}). 
In the case of afterglows seen off-axis, the magnetic fields normal to the blastwave become more important \citep[e.g.][]{2020MNRAS.491.5815G}. Because the observed polarisation mostly depends on the geometry, as the blastwave decelerates, the polarisation is predicted to change with time, in a way that depends on the viewing angle, the jet opening angle (which also captures the Lorentz factor), the emission structure of the jet and the presence of ordered fields \citep[see e.g.][]{2020ApJ...892..131S, 2021MNRAS.507.5340T, 2004ASPC..312..460R}. The closer we are observing to the jet axis, and/or the longer away from the jet break time (around which the polarization is expected to peak) we are observing, the lower the linear polarisation is expected to be.
It is therefore obvious that several intrinsic polarization values in Figure \ref{fig:pol_calc} are unrealistic. 
This figure is, however, model-free and has the aim to visualize the different possible contributions of a jet component at 7.2 days, for different intrinsic polarizations. 
This may be useful because the exact contribution of the non-thermal component (which was dominant in the first days) is difficult to constrain in the optical regime, based on the light curve decomposition or SED modeling alone \citep[][]{Andreoni2022,Pasham2022}.
For instance, Figure \ref{fig:pol_calc} shows that a jet polarized at the 10\% level (not very different from the measurements obtained for the previous relativistic TDEs), cannot contribute more than 30\% than the thermal component at 7.2 days.

\citet{Pasham2022} also suggested that the high energy emission (X-ray) originates from inverse Compton scattering (either synchrotron self-Compton or external Compton). Inverse-Compton is expected to produce non-intrinsic circular polarization in the case of a jet that is not made of pure electron-positron plasma, but also contains some fraction of protons \citep{Liodakis2022}. The degree of circular polarization depends on the strength of the magnetic field, the observing frequency, the uniformity of the magnetic field (which can be measured through the degree of linear polarization), the fraction of positrons and the Lorenz factor \citep[see eq. 1 in][]{Liodakis2022}. However, from the non-detection of significant circular and linear polarization in the case of AT~2022cmc, the ratio $p_{cir} / p_{lin}$ diverges and we cannot make any conclusions on the strength of the magnetic field and the fraction of positrons. 
We note that radio polarimetry would be more relevant for studies of jet properties in TDEs, compared to optical polarimetry. \citet{2003ApJ...595L..25M} calculated that in the presence of an ordered magnetic field, the intrinsic circular polarization of synchrotron emission can reach 1\% for forward shocks and 10\%-1\% for reverse shocks at radio frequencies, in contrast to 0.01\% and 0.1\%-0.01\% at optical frequencies, respectively \citep[see also][]{Wiersema2012}.
At face value, however, the polarization measurements seem compatible with the modeling of \citet{Pasham2022}, who favor a matter-dominated jet with low magnetic field energy density.
The non-detection of circular polarization may also imply no anisotropic pitch-angle distribution, which may produce circular polarization also in the case of random magnetic field orientations \citep[see e.g.][]{Wiersema2014}. 
Furthermore, in forward shocks, ordered magnetic fields originating from the central engine are not likely \citep{2003ApJ...595L..25M,2004ApJ...615..366S, Mundell2013, Wiersema2014}, which is also consistent with the non-detection of circular polarization. 

Therefore, the low observed linear and circular polarization degrees are compatible with the suggested explanations by \cite{Andreoni2022} and \cite{Pasham2022} that AT~2022cmc is a relativistic TDE seen pole on, that the optical/UV is dominated by a thermal component during the plateau phase, and that the jet is possibly matter-dominated.

\section{Summary and conclusions}
\label{sect:conc}

Linear and circular polarimetry of AT 2022cmc was performed with VLT/FORS2 in the $R$-band during the plateau phase, on February 27, 2022 and March 10, 2022 (7.22 and 12.23 rest-frame days after the detection), respectively. No obvious linear nor circular polarization was detected. The linear polarization degree is $p_{lin}$ = 0.14 $\pm$ 0.73 \%, with a 3$\sigma$ upper limit of 2.3 \%, and the circular polarization degree is $p_{cir}$ = -0.30 $\pm$ 0.53\%. 

The non-detection of polarization is consistent with the scenario that AT 2022cmc is a relativistic TDE and that, at the phases obtained, the observed emission most likely originates from a thermal component from the TDE, that is axially symmetric and is viewed pole-on. 

Next-generation time-domain surveys  (e.g. the Vera Rubin Observatory) will allow us to detect a large sample of TDEs and other transients in the near future. Our work demonstrates that it is important to conduct spectropolarimetric observations of the fast fading optical components of new (jetted) TDEs in real time, as this can help us probe the structure of the gas flow and constrain the origin of TDE emissions \citep{Roth20review, Dai21review}.\\

This work is based on observations collected at the European Organisation for Astronomical Research in the Southern Hemisphere under ESO programme 108.222Q.001 (PI Leloudas); the execution in service mode of these observations by the VLT operations staff is gratefully acknowledged. The work of A.C. is supported by NOIRLab, which is managed by the Association of Universities for Research in Astronomy (AURA) under a cooperative agreement with the National Science Foundation.
G.L. was supported by a research grant (19054) from VILLUM FONDEN.
M.B. acknowledges support from the European Union’s Horizon 2020 Programme under the AHEAD2020 project (grant agreement n. 871158). L.D. acknowledges the support from the Hong Kong Research Grants Council (HKU27305119, HKU17304821) and the National Natural Science Foundation of China (HKU12122309). We thank the anonymous referee for constructive comments.

\bibliography{TDEbibliography}{}

\begin{thebibliography}{}
\expandafter\ifx\csname natexlab\endcsname\relax\def\natexlab#1{#1}\fi
\providecommand{\url}[1]{\href{#1}{#1}}
\providecommand{\dodoi}[1]{doi:~\href{http://doi.org/#1}{\nolinkurl{#1}}}
\providecommand{\doeprint}[1]{\href{http://ascl.net/#1}{\nolinkurl{http://ascl.net/#1}}}
\providecommand{\doarXiv}[1]{\href{https://arxiv.org/abs/#1}{\nolinkurl{https://arxiv.org/abs/#1}}}

\bibitem[{{Andreoni} {et~al.}(2022{\natexlab{a}}){Andreoni}, {Coughlin},
  {Perley}, {Yao}, {Lu}, {Cenko}, {Kumar}, {Anand}, {Ho}, {Kasliwal}, {de
  Ugarte Postigo}, {Sagu{\'e}s-Carracedo}, {Schulze}, {Kann}, {Kulkarni},
  {Sollerman}, {Tanvir}, {Rest}, {Izzo}, {Somalwar}, {Kaplan}, {Ahumada},
  {Anupama}, {Auchettl}, {Barway}, {Bellm}, {Bhalerao}, {Bloom}, {Bremer},
  {Bulla}, {Burns}, {Campana}, {Chandra}, {Charalampopoulos}, {Cooke},
  {D'Elia}, {Das}, {Dobie}, {Fern{\'a}ndez}, {Freeburn}, {Fremling}, {Gezari},
  {Goode}, {Graham}, {Hammerstein}, {Karambelkar}, {Kilpatrick}, {Kool},
  {Krips}, {Laher}, {Leloudas}, {Levan}, {Lundquist}, {Mahabal}, {Medford},
  {Miller}, {M{\"o}ller}, {Mooley}, {Nayana}, {Nir}, {Pang}, {Paraskeva},
  {Perley}, {Petitpas}, {Pursiainen}, {Ravi}, {Ridden-Harper}, {Riddle},
  {Rigault}, {Rodriguez}, {Rusholme}, {Sharma}, {Smith}, {Stein}, {Th{\"o}ne},
  {Tohuvavohu}, {Valdes}, {van Roestel}, {Vergani}, {Wang}, \&
  {Zhang}}]{Andreoni2022}
{Andreoni}, I., {Coughlin}, M.~W., {Perley}, D.~A., {et~al.}
  2022{\natexlab{a}}, \nat, 612, 430, \dodoi{10.1038/s41586-022-05465-8}

\bibitem[{{Andreoni} {et~al.}(2022{\natexlab{b}}){Andreoni}, {Coughlin},
  {Ahumada}, {Kasliwal}, {Perley}, {Burns}, {Bulla}, {Cenko}, {Anand}, \&
  {Kool}}]{2022TNSAN..38....1A}
{Andreoni}, I., {Coughlin}, M., {Ahumada}, T., {et~al.} 2022{\natexlab{b}},
  Transient Name Server AstroNote, 38, 1

\bibitem[{{Appenzeller} {et~al.}(1998){Appenzeller}, {Fricke}, {F{\"u}rtig},
  {G{\"a}ssler}, {H{\"a}fner}, {Harke}, {Hess}, {Hummel}, {J{\"u}rgens},
  {Kudritzki}, {Mantel}, {Meisl}, {Muschielok}, {Nicklas}, {Rupprecht},
  {Seifert}, {Stahl}, {Szeifert}, \& {Tarantik}}]{1998Msngr..94....1A}
{Appenzeller}, I., {Fricke}, K., {F{\"u}rtig}, W., {et~al.} 1998, The
  Messenger, 94, 1

\bibitem[{{Arcavi} {et~al.}(2014){Arcavi}, {Gal-Yam}, {Sullivan}, {Pan},
  {Cenko}, {Horesh}, {Ofek}, {De Cia}, {Yan}, {Yang}, {Howell}, {Tal},
  {Kulkarni}, {Tendulkar}, {Tang}, {Xu}, {Sternberg}, {Cohen}, {Bloom},
  {Nugent}, {Kasliwal}, {Perley}, {Quimby}, {Miller}, {Theissen}, \&
  {Laher}}]{Arcavi2014}
{Arcavi}, I., {Gal-Yam}, A., {Sullivan}, M., {et~al.} 2014, \apj, 793, 38,
  \dodoi{10.1088/0004-637X/793/1/38}

\bibitem[{{Bagnulo} {et~al.}(2009){Bagnulo}, {Landolfi}, {Landstreet}, {Landi
  Degl'Innocenti}, {Fossati}, \& {Sterzik}}]{2009PASP..121..993B}
{Bagnulo}, S., {Landolfi}, M., {Landstreet}, J.~D., {et~al.} 2009, \pasp, 121,
  993, \dodoi{10.1086/605654}

\bibitem[{{Bellm} {et~al.}(2019){Bellm}, {Kulkarni}, {Graham}, {Dekany},
  {Smith}, {Riddle}, {Masci}, {Helou}, {Prince}, {Adams}, {Barbarino},
  {Barlow}, {Bauer}, {Beck}, {Belicki}, {Biswas}, {Blagorodnova}, {Bodewits},
  {Bolin}, {Brinnel}, {Brooke}, {Bue}, {Bulla}, {Burruss}, {Cenko}, {Chang},
  {Connolly}, {Coughlin}, {Cromer}, {Cunningham}, {De}, {Delacroix}, {Desai},
  {Duev}, {Eadie}, {Farnham}, {Feeney}, {Feindt}, {Flynn}, {Franckowiak},
  {Frederick}, {Fremling}, {Gal-Yam}, {Gezari}, {Giomi}, {Goldstein},
  {Golkhou}, {Goobar}, {Groom}, {Hacopians}, {Hale}, {Henning}, {Ho}, {Hover},
  {Howell}, {Hung}, {Huppenkothen}, {Imel}, {Ip}, {Ivezi{\'c}}, {Jackson},
  {Jones}, {Juric}, {Kasliwal}, {Kaspi}, {Kaye}, {Kelley}, {Kowalski},
  {Kramer}, {Kupfer}, {Landry}, {Laher}, {Lee}, {Lin}, {Lin}, {Lunnan},
  {Giomi}, {Mahabal}, {Mao}, {Miller}, {Monkewitz}, {Murphy}, {Ngeow},
  {Nordin}, {Nugent}, {Ofek}, {Patterson}, {Penprase}, {Porter}, {Rauch},
  {Rebbapragada}, {Reiley}, {Rigault}, {Rodriguez}, {van Roestel}, {Rusholme},
  {van Santen}, {Schulze}, {Shupe}, {Singer}, {Soumagnac}, {Stein}, {Surace},
  {Sollerman}, {Szkody}, {Taddia}, {Terek}, {Van Sistine}, {van Velzen},
  {Vestrand}, {Walters}, {Ward}, {Ye}, {Yu}, {Yan}, \& {Zolkower}}]{BellmZTF}
{Bellm}, E.~C., {Kulkarni}, S.~R., {Graham}, M.~J., {et~al.} 2019, \pasp, 131,
  018002, \dodoi{10.1088/1538-3873/aaecbe}

\bibitem[{{Blanchard} {et~al.}(2017){Blanchard}, {Nicholl}, {Berger},
  {Guillochon}, {Margutti}, {Chornock}, {Alexander}, {Leja}, \&
  {Drout}}]{BlanchardPS16dtm}
{Blanchard}, P.~K., {Nicholl}, M., {Berger}, E., {et~al.} 2017, \apj, 843, 106,
  \dodoi{10.3847/1538-4357/aa77f7}

\bibitem[{{Bloom} {et~al.}(2011){Bloom}, {Giannios}, {Metzger}, {Cenko},
  {Perley}, {Butler}, {Tanvir}, {Levan}, {O'Brien}, {Strubbe}, {De Colle},
  {Ramirez-Ruiz}, {Lee}, {Nayakshin}, {Quataert}, {King}, {Cucchiara},
  {Guillochon}, {Bower}, {Fruchter}, {Morgan}, \& {van der Horst}}]{Bloom2011}
{Bloom}, J.~S., {Giannios}, D., {Metzger}, B.~D., {et~al.} 2011, Science, 333,
  203, \dodoi{10.1126/science.1207150}

\bibitem[{{Brown} {et~al.}(2015){Brown}, {Levan}, {Stanway}, {Tanvir}, {Cenko},
  {Berger}, {Chornock}, \& {Cucchiaria}}]{Brown2015}
{Brown}, G.~C., {Levan}, A.~J., {Stanway}, E.~R., {et~al.} 2015, \mnras, 452,
  4297, \dodoi{10.1093/mnras/stv1520}

\bibitem[{{Bulla}(2019)}]{BullaPOSSIS}
{Bulla}, M. 2019, \mnras, 489, 5037, \dodoi{10.1093/mnras/stz2495}

\bibitem[{{Bulla} {et~al.}(2019){Bulla}, {Covino}, {Kyutoku}, {Tanaka},
  {Maund}, {Patat}, {Toma}, {Wiersema}, {Bruten}, {Jin}, \& {Testa}}]{BullaKN}
{Bulla}, M., {Covino}, S., {Kyutoku}, K., {et~al.} 2019, Nature Astronomy, 3,
  99, \dodoi{10.1038/s41550-018-0593-y}

\bibitem[{{Bulla} {et~al.}(2021){Bulla}, {Kyutoku}, {Tanaka}, {Covino},
  {Bruten}, {Matsumoto}, {Maund}, {Testa}, \& {Wiersema}}]{Bulla2021}
{Bulla}, M., {Kyutoku}, K., {Tanaka}, M., {et~al.} 2021, \mnras, 501, 1891,
  \dodoi{10.1093/mnras/staa3796}

\bibitem[{{Cenko} {et~al.}(2012){Cenko}, {Krimm}, {Horesh}, {Rau}, {Frail},
  {Kennea}, {Levan}, {Holland}, {Butler}, {Quimby}, {Bloom}, {Filippenko},
  {Gal-Yam}, {Greiner}, {Kulkarni}, {Ofek}, {Olivares E.}, {Schady},
  {Silverman}, {Tanvir}, \& {Xu}}]{Cenko2012}
{Cenko}, S.~B., {Krimm}, H.~A., {Horesh}, A., {et~al.} 2012, \apj, 753, 77,
  \dodoi{10.1088/0004-637X/753/1/77}

\bibitem[{{Charalampopoulos} {et~al.}(2022){Charalampopoulos}, {Bulla},
  {Bonnerot}, \& {Leloudas}}]{Charalampopoulos2022}
{Charalampopoulos}, P., {Bulla}, M., {Bonnerot}, C., \& {Leloudas}, G. 2022,
  arXiv e-prints, arXiv:2212.05079.
\newblock \doarXiv{2212.05079}

\bibitem[{{Chu} {et~al.}(2022){Chu}, {Cikota}, {Baade}, {Patat}, {Filippenko},
  {Wheeler}, {Maund}, {Bulla}, {Yang}, {H{\"o}flich}, \&
  {Wang}}]{2022MNRAS.509.6028C}
{Chu}, M.~R., {Cikota}, A., {Baade}, D., {et~al.} 2022, \mnras, 509, 6028,
  \dodoi{10.1093/mnras/stab3392}

\bibitem[{{Cikota} {et~al.}(2017){Cikota}, {Patat}, {Cikota}, \&
  {Faran}}]{2017MNRAS.464.4146C}
{Cikota}, A., {Patat}, F., {Cikota}, S., \& {Faran}, T. 2017, \mnras, 464,
  4146, \dodoi{10.1093/mnras/stw2545}

\bibitem[{{Cikota} {et~al.}(2018){Cikota}, {Leloudas}, {Bulla}, {Inserra},
  {Chen}, {Spyromilio}, {Patat}, {Cano}, {Cikota}, {Coughlin}, {Kankare},
  {Lowe}, {Maund}, {Rest}, {Smartt}, {Smith}, {Wainscoat}, \&
  {Young}}]{Cikota2018}
{Cikota}, A., {Leloudas}, G., {Bulla}, M., {et~al.} 2018, \mnras, 479, 4984,
  \dodoi{10.1093/mnras/sty1891}

\bibitem[{{Cikota} {et~al.}(2019){Cikota}, {Patat}, {Wang}, {Wheeler}, {Bulla},
  {Baade}, {H{\"o}flich}, {Cikota}, {Clocchiatti}, {Maund}, {Stevance}, \&
  {Yang}}]{2019MNRAS.490..578C}
{Cikota}, A., {Patat}, F., {Wang}, L., {et~al.} 2019, \mnras, 490, 578,
  \dodoi{10.1093/mnras/stz2322}

\bibitem[{{Coughlin} \& {Begelman}(2014)}]{2014ApJ...781...82C}
{Coughlin}, E.~R., \& {Begelman}, M.~C. 2014, \apj, 781, 82,
  \dodoi{10.1088/0004-637X/781/2/82}

\bibitem[{{Covino} \& {Gotz}(2016)}]{CovinoGotz}
{Covino}, S., \& {Gotz}, D. 2016, Astronomical and Astrophysical Transactions,
  29, 205.
\newblock \doarXiv{1605.03588}

\bibitem[{{Covino} {et~al.}(2017){Covino}, {Wiersema}, {Fan}, {Toma},
  {Higgins}, {Melandri}, {D'Avanzo}, {Mundell}, {Palazzi}, {Tanvir},
  {Bernardini}, {Branchesi}, {Brocato}, {Campana}, {di Serego Alighieri},
  {G{\"o}tz}, {Fynbo}, {Gao}, {Gomboc}, {Gompertz}, {Greiner}, {Hjorth}, {Jin},
  {Kaper}, {Klose}, {Kobayashi}, {Kopac}, {Kouveliotou}, {Levan}, {Mao},
  {Malesani}, {Pian}, {Rossi}, {Salvaterra}, {Starling}, {Steele},
  {Tagliaferri}, {Troja}, {van der Horst}, \& {Wijers}}]{Covino2017}
{Covino}, S., {Wiersema}, K., {Fan}, Y.~Z., {et~al.} 2017, Nature Astronomy, 1,
  791, \dodoi{10.1038/s41550-017-0285-z}

\bibitem[{{Dai} {et~al.}(2021){Dai}, {Lodato}, \& {Cheng}}]{Dai21review}
{Dai}, J.~L., {Lodato}, G., \& {Cheng}, R. 2021, \ssr, 217, 12,
  \dodoi{10.1007/s11214-020-00747-x}

\bibitem[{{Dai} {et~al.}(2018){Dai}, {McKinney}, {Roth}, {Ramirez-Ruiz}, \&
  {Miller}}]{Dai18}
{Dai}, L., {McKinney}, J.~C., {Roth}, N., {Ramirez-Ruiz}, E., \& {Miller},
  M.~C. 2018, \apj, 859, L20, \dodoi{10.3847/2041-8213/aab429}

\bibitem[{{Dong} {et~al.}(2016){Dong}, {Shappee}, {Prieto}, {Jha}, {Stanek},
  {Holoien}, {Kochanek}, {Thompson}, {Morrell}, {Thompson}, {Basu}, {Beacom},
  {Bersier}, {Brimacombe}, {Brown}, {Bufano}, {Chen}, {Conseil}, {Danilet},
  {Falco}, {Grupe}, {Kiyota}, {Masi}, {Nicholls}, {Olivares E.}, {Pignata},
  {Pojmanski}, {Simonian}, {Szczygiel}, \& {Wo{\'z}niak}}]{2016Sci...351..257D}
{Dong}, S., {Shappee}, B.~J., {Prieto}, J.~L., {et~al.} 2016, Science, 351,
  257, \dodoi{10.1126/science.aac9613}

\bibitem[{ESO(2015)}]{FORS2manual}
ESO. 2015, FORS2 User Manual, Vol. 96.0 (European Southern Observatory)

\bibitem[{{Eyles-Ferris} {et~al.}(2022){Eyles-Ferris}, {Starling}, {O'Brien},
  {Nixon}, \& {Coughlin}}]{2022MNRAS.517.6013E}
{Eyles-Ferris}, R.~A.~J., {Starling}, R.~L.~C., {O'Brien}, P.~T., {Nixon},
  C.~J., \& {Coughlin}, E.~R. 2022, \mnras, 517, 6013,
  \dodoi{10.1093/mnras/stac3073}

\bibitem[{{French} {et~al.}(2020){French}, {Wevers}, {Law-Smith}, {Graur}, \&
  {Zabludoff}}]{french2020ssrv}
{French}, K.~D., {Wevers}, T., {Law-Smith}, J., {Graur}, O., \& {Zabludoff},
  A.~I. 2020, \ssr, 216, 32, \dodoi{10.1007/s11214-020-00657-y}

\bibitem[{{Gezari} {et~al.}(2006){Gezari}, {Martin}, {Milliard}, {Basa},
  {Halpern}, {Forster}, {Friedman}, {Morrissey}, {Neff}, {Schiminovich},
  {Seibert}, {Small}, \& {Wyder}}]{Gezari2006}
{Gezari}, S., {Martin}, D.~C., {Milliard}, B., {et~al.} 2006, \apjl, 653, L25,
  \dodoi{10.1086/509918}

\bibitem[{{Gezari} {et~al.}(2012){Gezari}, {Chornock}, {Rest}, {Huber},
  {Forster}, {Berger}, {Challis}, {Neill}, {Martin}, {Heckman}, {Lawrence},
  {Norman}, {Narayan}, {Foley}, {Marion}, {Scolnic}, {Chomiuk}, {Soderberg},
  {Smith}, {Kirshner}, {Riess}, {Smartt}, {Stubbs}, {Tonry}, {Wood-Vasey},
  {Burgett}, {Chambers}, {Grav}, {Heasley}, {Kaiser}, {Kudritzki}, {Magnier},
  {Morgan}, \& {Price}}]{Gezari2012}
{Gezari}, S., {Chornock}, R., {Rest}, A., {et~al.} 2012, \nat, 485, 217,
  \dodoi{10.1038/nature10990}

\bibitem[{{Gill} \& {Granot}(2020)}]{2020MNRAS.491.5815G}
{Gill}, R., \& {Granot}, J. 2020, \mnras, 491, 5815,
  \dodoi{10.1093/mnras/stz3340}

\bibitem[{{Gonz{\'a}lez-Gait{\'a}n} {et~al.}(2020){Gonz{\'a}lez-Gait{\'a}n},
  {Mour{\~a}o}, {Patat}, {Anderson}, {Cikota}, {Wiersema}, {Higgins}, \&
  {Silva}}]{2020A&A...634A..70G}
{Gonz{\'a}lez-Gait{\'a}n}, S., {Mour{\~a}o}, A.~M., {Patat}, F., {et~al.} 2020,
  \aap, 634, A70, \dodoi{10.1051/0004-6361/201936379}

\bibitem[{{Granot}(2003)}]{2003ApJ...596L..17G}
{Granot}, J. 2003, \apjl, 596, L17, \dodoi{10.1086/379110}

\bibitem[{{Granot} \& {K{\"o}nigl}(2003)}]{2003ApJ...594L..83G}
{Granot}, J., \& {K{\"o}nigl}, A. 2003, \apjl, 594, L83, \dodoi{10.1086/378733}

\bibitem[{{Granot} \& {Ramirez-Ruiz}(2010)}]{2010arXiv1012.5101G}
{Granot}, J., \& {Ramirez-Ruiz}, E. 2010, arXiv e-prints, arXiv:1012.5101.
\newblock \doarXiv{1012.5101}

\bibitem[{{Heiles}(2000)}]{2000AJ....119..923H}
{Heiles}, C. 2000, \aj, 119, 923, \dodoi{10.1086/301236}

\bibitem[{{Higgins} {et~al.}(2019){Higgins}, {Wiersema}, {Covino}, {Starling},
  {Stevance}, {Wyrzykowski}, {Hodgkin}, {Maund}, {O'Brien}, \&
  {Tanvir}}]{HigginsSPLOT}
{Higgins}, A.~B., {Wiersema}, K., {Covino}, S., {et~al.} 2019, \mnras, 482,
  5023, \dodoi{10.1093/mnras/sty3029}

\bibitem[{{Hoflich}(1991)}]{Hoeflich1991}
{Hoflich}, P. 1991, \aap, 246, 481

\bibitem[{{Holoien} {et~al.}(2020){Holoien}, {Auchettl}, {Tucker}, {Shappee},
  {Patel}, {Miller-Jones}, {Mockler}, {Groenewald}, {Hinkle}, {Brown},
  {Kochanek}, {Stanek}, {Chen}, {Dong}, {Prieto}, {Thompson}, {Beaton},
  {Connor}, {Cowperthwaite}, {Dahmen}, {French}, {Morrell}, {Buckley},
  {Gromadzki}, {Roy}, {Coulter}, {Dimitriadis}, {Foley}, {Kilpatrick}, {Piro},
  {Rojas-Bravo}, {Siebert}, \& {Velzen}}]{Holoien2020}
{Holoien}, T. W.~S., {Auchettl}, K., {Tucker}, M.~A., {et~al.} 2020, \apj, 898,
  161, \dodoi{10.3847/1538-4357/ab9f3d}

\bibitem[{{Hutsem{\'e}kers} {et~al.}(2010){Hutsem{\'e}kers}, {Borguet},
  {Sluse}, {Cabanac}, \& {Lamy}}]{2010A&A...520L...7H}
{Hutsem{\'e}kers}, D., {Borguet}, B., {Sluse}, D., {Cabanac}, R., \& {Lamy}, H.
  2010, \aap, 520, L7, \dodoi{10.1051/0004-6361/201015359}

\bibitem[{{Kasen} {et~al.}(2003){Kasen}, {Nugent}, {Wang}, {Howell}, {Wheeler},
  {H{\"o}flich}, {Baade}, {Baron}, \& {Hauschildt}}]{Kasen2003}
{Kasen}, D., {Nugent}, P., {Wang}, L., {et~al.} 2003, \apj, 593, 788,
  \dodoi{10.1086/376601}

\bibitem[{{Komossa} \& {Bade}(1999)}]{komossa99}
{Komossa}, S., \& {Bade}, N. 1999, \aap, 343, 775.
\newblock \doarXiv{astro-ph/9901141}

\bibitem[{{Kremer} {et~al.}(2021){Kremer}, {Lu}, {Piro}, {Chatterjee}, {Rasio},
  \& {Ye}}]{2021ApJ...911..104K}
{Kremer}, K., {Lu}, W., {Piro}, A.~L., {et~al.} 2021, \apj, 911, 104,
  \dodoi{10.3847/1538-4357/abeb14}

\bibitem[{{Kuin} {et~al.}(2019){Kuin}, {Wu}, {Oates}, {Lien}, {Emery},
  {Kennea}, {de Pasquale}, {Han}, {Brown}, {Tohuvavohu}, {Breeveld}, {Burrows},
  {Cenko}, {Campana}, {Levan}, {Markwardt}, {Osborne}, {Page}, {Page},
  {Sbarufatti}, {Siegel}, \& {Troja}}]{2019MNRAS.487.2505K}
{Kuin}, N. P.~M., {Wu}, K., {Oates}, S., {et~al.} 2019, \mnras, 487, 2505,
  \dodoi{10.1093/mnras/stz053}

\bibitem[{{Lazzati}(2006)}]{2006NJPh....8..131L}
{Lazzati}, D. 2006, New Journal of Physics, 8, 131,
  \dodoi{10.1088/1367-2630/8/8/131}

\bibitem[{{Lee} {et~al.}(2020){Lee}, {Hung}, {Matheson}, {Soraisam}, {Narayan},
  {Saha}, {Stubens}, \& {Wolf}}]{Lee2020}
{Lee}, C.-H., {Hung}, T., {Matheson}, T., {et~al.} 2020, \apjl, 892, L1,
  \dodoi{10.3847/2041-8213/ab7cd3}

\bibitem[{{Leloudas} {et~al.}(2015){Leloudas}, {Patat}, {Maund}, {Hsiao},
  {Malesani}, {Schulze}, {Contreras}, {de Ugarte Postigo}, {Sollerman},
  {Stritzinger}, {Taddia}, {Wheeler}, \& {Gorosabel}}]{2015ApJ...815L..10L}
{Leloudas}, G., {Patat}, F., {Maund}, J.~R., {et~al.} 2015, \apjl, 815, L10,
  \dodoi{10.1088/2041-8205/815/1/L10}

\bibitem[{{Leloudas} {et~al.}(2016){Leloudas}, {Fraser}, {Stone}, {van Velzen},
  {Jonker}, {Arcavi}, {Fremling}, {Maund}, {Smartt}, {Kr{\`\i}hler},
  {Miller-Jones}, {Vreeswijk}, {Gal-Yam}, {Mazzali}, {De Cia}, {Howell},
  {Inserra}, {Patat}, {de Ugarte Postigo}, {Yaron}, {Ashall}, {Bar},
  {Campbell}, {Chen}, {Childress}, {Elias-Rosa}, {Harmanen}, {Hosseinzadeh},
  {Johansson}, {Kangas}, {Kankare}, {Kim}, {Kuncarayakti}, {Lyman}, {Magee},
  {Maguire}, {Malesani}, {Mattila}, {McCully}, {Nicholl}, {Prentice},
  {Romero-Ca{\~n}izales}, {Schulze}, {Smith}, {Sollerman}, {Sullivan},
  {Tucker}, {Valenti}, {Wheeler}, \& {Young}}]{Leloudas2016}
{Leloudas}, G., {Fraser}, M., {Stone}, N.~C., {et~al.} 2016, Nature Astronomy,
  1, 0002, \dodoi{10.1038/s41550-016-0002}

\bibitem[{{Leloudas} {et~al.}(2022){Leloudas}, {Bulla}, {Cikota}, {Dai},
  {Thomsen}, {Maund}, {Charalampopoulos}, {Roth}, {Arcavi}, {Auchettl},
  {Malesani}, {Nicholl}, \& {Ramirez-Ruiz}}]{Leloudas2022}
{Leloudas}, G., {Bulla}, M., {Cikota}, A., {et~al.} 2022, Nature Astronomy, 6,
  1193, \dodoi{10.1038/s41550-022-01767-z}

\bibitem[{{Levan} {et~al.}(2011){Levan}, {Tanvir}, {Cenko}, {Perley},
  {Wiersema}, {Bloom}, {Fruchter}, {Postigo}, {O'Brien}, {Butler}, {van der
  Horst}, {Leloudas}, {Morgan}, {Misra}, {Bower}, {Farihi}, {Tunnicliffe},
  {Modjaz}, {Silverman}, {Hjorth}, {Th{\"o}ne}, {Cucchiara}, {Cer{\'o}n},
  {Castro-Tirado}, {Arnold}, {Bremer}, {Brodie}, {Carroll}, {Cooper}, {Curran},
  {Cutri}, {Ehle}, {Forbes}, {Fynbo}, {Gorosabel}, {Graham}, {Hoffman},
  {Guziy}, {Jakobsson}, {Kamble}, {Kerr}, {Kasliwal}, {Kouveliotou},
  {Kocevski}, {Law}, {Nugent}, {Ofek}, {Poznanski}, {Quimby}, {Rol},
  {Romanowsky}, {S{\'a}nchez-Ram{\'{\i}}rez}, {Schulze}, {Singh}, {van
  Spaandonk}, {Starling}, {Strom}, {Tello}, {Vaduvescu}, {Wheatley}, {Wijers},
  {Winters}, \& {Xu}}]{Levan2011}
{Levan}, A.~J., {Tanvir}, N.~R., {Cenko}, S.~B., {et~al.} 2011, Science, 333,
  199, \dodoi{10.1126/science.1207143}

\bibitem[{{Liodakis} {et~al.}(2022{\natexlab{a}}){Liodakis}, {Blinov},
  {Potter}, \& {Rieger}}]{Liodakis2022}
{Liodakis}, I., {Blinov}, D., {Potter}, S.~B., \& {Rieger}, F.~M.
  2022{\natexlab{a}}, \mnras, 509, L21, \dodoi{10.1093/mnrasl/slab118}

\bibitem[{{Liodakis} {et~al.}(2022{\natexlab{b}}){Liodakis}, {Koljonen},
  {Blinov}, {Lindfors}, {Alexander}, {Hovatta}, {Berton}, {Hajela},
  {Jormanainen}, {Kouroumpatzakis}, {Mandarakas}, \&
  {Nilsson}}]{2022arXiv220814465L}
{Liodakis}, I., {Koljonen}, K.~I.~I., {Blinov}, D., {et~al.}
  2022{\natexlab{b}}, arXiv e-prints, arXiv:2208.14465.
\newblock \doarXiv{2208.14465}

\bibitem[{{Liu} {et~al.}(2022){Liu}, {Zhu}, {Liu}, {Yu}, \&
  {Zhang}}]{2022ApJ...935L..34L}
{Liu}, J.-F., {Zhu}, J.-P., {Liu}, L.-D., {Yu}, Y.-W., \& {Zhang}, B. 2022,
  \apjl, 935, L34, \dodoi{10.3847/2041-8213/ac86d2}

\bibitem[{{Lu} \& {Bonnerot}(2020)}]{LuBonnerot2020}
{Lu}, W., \& {Bonnerot}, C. 2020, \mnras, 492, 686,
  \dodoi{10.1093/mnras/stz3405}

\bibitem[{{Marin} \& {Stalevski}(2015)}]{2015sf2a.conf..167M}
{Marin}, F., \& {Stalevski}, M. 2015, in SF2A-2015: Proceedings of the Annual
  meeting of the French Society of Astronomy and Astrophysics, 167--170.
\newblock \doarXiv{1510.01059}

\bibitem[{{Matsumiya} \& {Ioka}(2003)}]{2003ApJ...595L..25M}
{Matsumiya}, M., \& {Ioka}, K. 2003, \apjl, 595, L25, \dodoi{10.1086/378879}

\bibitem[{{Mattila} {et~al.}(2018){Mattila}, {P{\'e}rez-Torres}, {Efstathiou},
  {Mimica}, {Fraser}, {Kankare}, {Alberdi}, {Aloy}, {Heikkil{\"a}}, {Jonker},
  {Lundqvist}, {Mart{\'\i}-Vidal}, {Meikle}, {Romero-Ca{\~n}izales}, {Smartt},
  {Tsygankov}, {Varenius}, {Alonso-Herrero}, {Bondi}, {Fransson},
  {Herrero-Illana}, {Kangas}, {Kotak}, {Ram{\'\i}rez-Olivencia},
  {V{\"a}is{\"a}nen}, {Beswick}, {Clements}, {Greimel}, {Harmanen},
  {Kotilainen}, {Nandra}, {Reynolds}, {Ryder}, {Walton}, {Wiik}, \&
  {{\"O}stlin}}]{Mattila2018}
{Mattila}, S., {P{\'e}rez-Torres}, M., {Efstathiou}, A., {et~al.} 2018,
  Science, 361, 482, \dodoi{10.1126/science.aao4669}

\bibitem[{{Maund} {et~al.}(2020){Maund}, {Leloudas}, {Malesani}, {Patat},
  {Sollerman}, \& {de Ugarte Postigo}}]{Maund2020}
{Maund}, J.~R., {Leloudas}, G., {Malesani}, D.~B., {et~al.} 2020, \mnras, 498,
  3730, \dodoi{10.1093/mnras/staa2517}

\bibitem[{{Maund} {et~al.}(2013){Maund}, {Spyromilio}, {Hoflich}, {Wheeler},
  {Baade}, {Clocchiatti}, {Patat}, {Reilly}, {Wang}, \&
  {Zelaya}}]{2013MNRAS.433L..20M}
{Maund}, J.~R., {Spyromilio}, J., {Hoflich}, P.~A., {et~al.} 2013, \mnras, 433,
  L20, \dodoi{10.1093/mnrasl/slt050}

\bibitem[{{Metzger}(2022)}]{2022ApJ...932...84M}
{Metzger}, B.~D. 2022, \apj, 932, 84, \dodoi{10.3847/1538-4357/ac6d59}

\bibitem[{{Mundell} {et~al.}(2013){Mundell}, {Kopa{\v{c}}}, {Arnold}, {Steele},
  {Gomboc}, {Kobayashi}, {Harrison}, {Smith}, {Guidorzi}, {Virgili},
  {Melandri}, \& {Japelj}}]{Mundell2013}
{Mundell}, C.~G., {Kopa{\v{c}}}, D., {Arnold}, D.~M., {et~al.} 2013, \nat, 504,
  119, \dodoi{10.1038/nature12814}

\bibitem[{{Nakar} {et~al.}(2003){Nakar}, {Piran}, \&
  {Waxman}}]{2003JCAP...10..005N}
{Nakar}, E., {Piran}, T., \& {Waxman}, E. 2003, \jcap, 2003, 005,
  \dodoi{10.1088/1475-7516/2003/10/005}

\bibitem[{{Pasham} {et~al.}(2015){Pasham}, {Cenko}, {Levan}, {Bower}, {Horesh},
  {Brown}, {Dolan}, {Wiersema}, {Filippenko}, {Fruchter}, {Greiner}, {O'Brien},
  {Page}, {Rau}, \& {Tanvir}}]{2015ApJ...805...68P}
{Pasham}, D.~R., {Cenko}, S.~B., {Levan}, A.~J., {et~al.} 2015, \apj, 805, 68,
  \dodoi{10.1088/0004-637X/805/1/68}

\bibitem[{{Pasham} {et~al.}(2022){Pasham}, {Lucchini}, {Laskar}, {Gompertz},
  {Srivastav}, {Nicholl}, {Smartt}, {Miller-Jones}, {Alexander}, {Fender},
  {Smith}, {Fulton}, {Dewangan}, {Gendreau}, {Coughlin}, {Rhodes}, {Horesh},
  {van Velzen}, {Sfaradi}, {Guolo}, {Castro Segura}, {Aamer}, {Anderson},
  {Arcavi}, {Brennan}, {Chambers}, {Charalampopoulos}, {Chen}, {Clocchiatti},
  {de Boer}, {Dennefeld}, {Ferrara}, {Galbany}, {Gao}, {Gillanders}, {Goodwin},
  {Gromadzki}, {Huber}, {Jonker}, {Joshi}, {Kara}, {Killestein}, {Kosec},
  {Kocevski}, {Leloudas}, {Lin}, {Margutti}, {Mattila}, {Moore},
  {Muller-Bravo}, {Ngeow}, {Oates}, {Onori}, {Pan}, {Perez-Torres}, {Rani},
  {Remillard}, {Ridley}, {Schulze}, {Sheng}, {Shingles}, {Smith}, {Steiner},
  {Wainscoat}, {Wevers}, \& {Yang}}]{Pasham2022}
{Pasham}, D.~R., {Lucchini}, M., {Laskar}, T., {et~al.} 2022, arXiv e-prints,
  arXiv:2211.16537.
\newblock \doarXiv{2211.16537}

\bibitem[{{Patat}(2017)}]{PatatSNhandbook}
{Patat}, F. 2017, in Handbook of Supernovae, ed. A.~W. {Alsabti} \&
  P.~{Murdin}, 1017, \dodoi{10.1007/978-3-319-21846-5\_110}

\bibitem[{{Patat} \& {Romaniello}(2006)}]{Patat2006}
{Patat}, F., \& {Romaniello}, M. 2006, \pasp, 118, 146, \dodoi{10.1086/497581}

\bibitem[{{Patra} {et~al.}(2022){Patra}, {Lu}, {Brink}, {Yang}, {Filippenko},
  \& {Vasylyev}}]{Patra2022}
{Patra}, K.~C., {Lu}, W., {Brink}, T.~G., {et~al.} 2022, \mnras, 515, 138,
  \dodoi{10.1093/mnras/stac1727}

\bibitem[{{Perley} {et~al.}(2019){Perley}, {Mazzali}, {Yan}, {Cenko}, {Gezari},
  {Taggart}, {Blagorodnova}, {Fremling}, {Mockler}, {Singh}, {Tominaga},
  {Tanaka}, {Watson}, {Ahumada}, {Anupama}, {Ashall}, {Becerra}, {Bersier},
  {Bhalerao}, {Bloom}, {Butler}, {Copperwheat}, {Coughlin}, {De}, {Drake},
  {Duev}, {Frederick}, {Gonz{\'a}lez}, {Goobar}, {Heida}, {Ho}, {Horst},
  {Hung}, {Itoh}, {Jencson}, {Kasliwal}, {Kawai}, {Khanam}, {Kulkarni},
  {Kumar}, {Kumar}, {Kutyrev}, {Lee}, {Maeda}, {Mahabal}, {Murata}, {Neill},
  {Ngeow}, {Penprase}, {Pian}, {Quimby}, {Ramirez-Ruiz}, {Richer},
  {Rom{\'a}n-Z{\'u}{\~n}iga}, {Sahu}, {Srivastav}, {Socia}, {Sollerman},
  {Tachibana}, {Taddia}, {Tinyanont}, {Troja}, {Ward}, {Wee}, \&
  {Yu}}]{2019MNRAS.484.1031P}
{Perley}, D.~A., {Mazzali}, P.~A., {Yan}, L., {et~al.} 2019, \mnras, 484, 1031,
  \dodoi{10.1093/mnras/sty3420}

\bibitem[{{Phinney}(1989)}]{Phinney1989}
{Phinney}, E.~S. 1989, in IAU Symposium, Vol. 136, The Center of the Galaxy,
  ed. M.~{Morris}, 543

\bibitem[{{Plaszczynski} {et~al.}(2014){Plaszczynski}, {Montier}, {Levrier}, \&
  {Tristram}}]{2014MNRAS.439.4048P}
{Plaszczynski}, S., {Montier}, L., {Levrier}, F., \& {Tristram}, M. 2014,
  \mnras, 439, 4048, \dodoi{10.1093/mnras/stu270}

\bibitem[{{Rees}(1988)}]{Rees1988}
{Rees}, M.~J. 1988, \nat, 333, 523, \dodoi{10.1038/333523a0}

\bibitem[{{Rossi} {et~al.}(2004){Rossi}, {Lazzati}, {Salmonson}, \&
  {Ghisellini}}]{2004ASPC..312..460R}
{Rossi}, E., {Lazzati}, D., {Salmonson}, J.~D., \& {Ghisellini}, G. 2004, in
  Astronomical Society of the Pacific Conference Series, Vol. 312, Gamma-Ray
  Bursts in the Afterglow Era, ed. M.~{Feroci}, F.~{Frontera}, N.~{Masetti}, \&
  L.~{Piro}, 460

\bibitem[{{Roth} {et~al.}(2020){Roth}, {Rossi}, {Krolik}, {Piran}, {Mockler},
  \& {Kasen}}]{Roth20review}
{Roth}, N., {Rossi}, E.~M., {Krolik}, J., {et~al.} 2020, \ssr, 216, 114,
  \dodoi{10.1007/s11214-020-00735-1}

\bibitem[{{Sagiv} {et~al.}(2004){Sagiv}, {Waxman}, \&
  {Loeb}}]{2004ApJ...615..366S}
{Sagiv}, A., {Waxman}, E., \& {Loeb}, A. 2004, \apj, 615, 366,
  \dodoi{10.1086/423977}

\bibitem[{{Sarazin} \& {Roddier}(1990)}]{1990A&A...227..294S}
{Sarazin}, M., \& {Roddier}, F. 1990, \aap, 227, 294

\bibitem[{{Saxton} {et~al.}(2020){Saxton}, {Komossa}, {Auchettl}, \&
  {Jonker}}]{2020SSRv..216...85S}
{Saxton}, R., {Komossa}, S., {Auchettl}, K., \& {Jonker}, P.~G. 2020, \ssr,
  216, 85, \dodoi{10.1007/s11214-020-00708-4}

\bibitem[{{Schlafly} \& {Finkbeiner}(2011)}]{2011ApJ...737..103S}
{Schlafly}, E.~F., \& {Finkbeiner}, D.~P. 2011, \apj, 737, 103,
  \dodoi{10.1088/0004-637X/737/2/103}

\bibitem[{{Stringer} \& {Lazzati}(2020)}]{2020ApJ...892..131S}
{Stringer}, E., \& {Lazzati}, D. 2020, \apj, 892, 131,
  \dodoi{10.3847/1538-4357/ab76d2}

\bibitem[{{Tadhunter} {et~al.}(2017){Tadhunter}, {Spence}, {Rose}, {Mullaney},
  \& {Crowther}}]{Tadhunter2017}
{Tadhunter}, C., {Spence}, R., {Rose}, M., {Mullaney}, J., \& {Crowther}, P.
  2017, Nature Astronomy, 1, 0061, \dodoi{10.1038/s41550-017-0061}

\bibitem[{{Tanvir} {et~al.}(2022){Tanvir}, {de Ugarte Postigo}, {Izzo},
  {Vergani}, {D'Elia}, {Campana}, {Perley}, {Wiersema}, {Levan}, {Kann},
  {Rossi}, {Della Valle}, \& {Stargate Consortium}}]{2022GCN.31602....1T}
{Tanvir}, N.~R., {de Ugarte Postigo}, A., {Izzo}, L., {et~al.} 2022, GRB
  Coordinates Network, 31602, 1

\bibitem[{{Teboul} \& {Shaviv}(2021)}]{2021MNRAS.507.5340T}
{Teboul}, O., \& {Shaviv}, N.~J. 2021, \mnras, 507, 5340,
  \dodoi{10.1093/mnras/stab2491}

\bibitem[{{Toma} {et~al.}(2008){Toma}, {Ioka}, \&
  {Nakamura}}]{2008ApJ...673L.123T}
{Toma}, K., {Ioka}, K., \& {Nakamura}, T. 2008, \apjl, 673, L123,
  \dodoi{10.1086/528740}

\bibitem[{{van Velzen} {et~al.}(2020){van Velzen}, {Holoien}, {Onori}, {Hung},
  \& {Arcavi}}]{vanvelzen2020}
{van Velzen}, S., {Holoien}, T. W.~S., {Onori}, F., {Hung}, T., \& {Arcavi}, I.
  2020, \ssr, 216, 124, \dodoi{10.1007/s11214-020-00753-z}

\bibitem[{{Wang} \& {Wheeler}(2008)}]{WangWheeler}
{Wang}, L., \& {Wheeler}, J.~C. 2008, \araa, 46, 433,
  \dodoi{10.1146/annurev.astro.46.060407.145139}

\bibitem[{{Wiersema} {et~al.}(2012){Wiersema}, {van der Horst}, {Levan},
  {Tanvir}, {Karjalainen}, {Kamble}, {Kouveliotou}, {Metzger}, {Russell},
  {Skillen}, {Starling}, \& {Wijers}}]{Wiersema2012}
{Wiersema}, K., {van der Horst}, A.~J., {Levan}, A.~J., {et~al.} 2012, \mnras,
  421, 1942, \dodoi{10.1111/j.1365-2966.2011.20379.x}

\bibitem[{{Wiersema} {et~al.}(2014){Wiersema}, {Covino}, {Toma}, {van der
  Horst}, {Varela}, {Min}, {Greiner}, {Starling}, {Tanvir}, {Wijers},
  {Campana}, {Curran}, {Fan}, {Fynbo}, {Gorosabel}, {Gomboc}, {G{\"o}tz},
  {Hjorth}, {Jin}, {Kobayashi}, {Kouveliotou}, {Mundell}, {O'Brien}, {Pian},
  {Rowlinson}, {Russell}, {Salvaterra}, {di Serego Alighieri}, {Tagliaferri},
  {Vergani}, {Elliott}, {Fari{\~n}a}, {Hartoog}, {Karjalainen}, {Klose},
  {Knust}, {Levan}, {Schady}, {Sudilovsky}, \& {Willingale}}]{Wiersema2014}
{Wiersema}, K., {Covino}, S., {Toma}, K., {et~al.} 2014, \nat, 509, 201,
  \dodoi{10.1038/nature13237}

\bibitem[{{Wiersema} {et~al.}(2020){Wiersema}, {Higgins}, {Levan}, {Eyles},
  {Starling}, {Tanvir}, {Cenko}, {van der Horst}, {Gompertz}, {Greiner}, \&
  {Pasham}}]{Wiersema2020}
{Wiersema}, K., {Higgins}, A.~B., {Levan}, A.~J., {et~al.} 2020, \mnras, 491,
  1771, \dodoi{10.1093/mnras/stz3106}

\bibitem[{{Zabludoff} {et~al.}(2021){Zabludoff}, {Arcavi}, {La Massa},
  {Perets}, {Trakhtenbrot}, {Zauderer}, {Auchettl}, {Dai}, {French}, {Hung},
  {Kara}, {Lodato}, {Maksym}, {Qin}, {Ramirez-Ruiz}, {Roth}, {Runnoe}, \&
  {Wevers}}]{Zabludoff2021}
{Zabludoff}, A., {Arcavi}, I., {La Massa}, S., {et~al.} 2021, \ssr, 217, 54,
  \dodoi{10.1007/s11214-021-00829-4}

\bibitem[{{Zauderer} {et~al.}(2011){Zauderer}, {Berger}, {Soderberg}, {Loeb},
  {Narayan}, {Frail}, {Petitpas}, {Brunthaler}, {Chornock}, {Carpenter},
  {Pooley}, {Mooley}, {Kulkarni}, {Margutti}, {Fox}, {Nakar}, {Patel},
  {Volgenau}, {Culverhouse}, {Bietenholz}, {Rupen}, {Max-Moerbeck}, {Readhead},
  {Richards}, {Shepherd}, {Storm}, \& {Hull}}]{Zauderer2011}
{Zauderer}, B.~A., {Berger}, E., {Soderberg}, A.~M., {et~al.} 2011, \nat, 476,
  425, \dodoi{10.1038/nature10366}

\end{thebibliography}
\bibliographystyle{aasjournal}

\end{document}